\pdfoutput=1

\documentclass[11pt,dvipsnames]{article}

\usepackage{acl}

\usepackage{times}
\usepackage{latexsym}
\usepackage{url}            %
\usepackage{wrapfig}
\usepackage{booktabs}       %

\usepackage{caption}
\usepackage{placeins}
\usepackage{amssymb}%
\usepackage{pifont}%
\usepackage{floatrow}
\usepackage{fancyvrb}
\usepackage{graphicx}
\usepackage{amsmath}
\usepackage{amsfonts}       %
\usepackage{nicefrac}       %
\usepackage{microtype}      %
\usepackage{xcolor}         %
\usepackage{xspace}         %
\usepackage{multirow}
\usepackage{array}
\usepackage{siunitx}
\usepackage{booktabs}
\usepackage{colortbl}
\usepackage{color}
\usepackage{soul}
\usepackage{highlight}
\usepackage{tcolorbox}
\usepackage{hyperref}

\def\Snospace~{\S{}}

\newcommand\extrafootertext[1]{%
    \bgroup
    \renewcommand\thefootnote{\fnsymbol{footnote}}%
    \renewcommand\thempfootnote{\fnsymbol{mpfootnote}}%
    \footnotetext[0]{#1}%
    \egroup
}

\usepackage[T1]{fontenc}
\definecolor{mygray}{RGB}{220, 220, 220}
\definecolor{lightgreen}{RGB}{223,255,219}
\definecolor{lightred}{RGB}{255,219,219}

\definecolor{pos}{RGB}{167, 199, 231}
\definecolor{neg}{RGB}{250, 160, 160}

\usepackage{soul}

\DeclareRobustCommand{\hlpos}[1]{{\sethlcolor{pos}\hl{#1}}}
\DeclareRobustCommand{\hlneg}[1]{{\sethlcolor{neg}\hl{#1}}}
\DeclareRobustCommand{\hlbase}[1]{{\sethlcolor{lightgray}\hl{#1}}}

\usepackage[utf8]{inputenc}

\newcommand{\chatgpt}{\texttt{gpt-3.5-turbo-0613}\xspace}
\newcommand{\chatgptShort}{\texttt{gpt-3.5-turbo}\xspace}

\newcommand{\gptfour}{\texttt{gpt-4-0613}\xspace}

\usepackage{microtype}

\title{When do Generative Query and Document Expansions Fail? \\ A Comprehensive Study Across Methods, Retrievers, and Datasets}

\author{
    \textbf{Orion Weller}$^{\hspace{.1em}{\color{BrickRed}\boldsymbol{\ast}}\hspace{.1em}{\color{RoyalBlue}\boldsymbol{\iota}}}$
    \quad
    \textbf{Kyle Lo}$^{\hspace{.1em}\color{RoyalBlue}\boldsymbol{\alpha}}$
    \quad
    \textbf{David Wadden}$^{\hspace{.1em}\color{RoyalBlue}\boldsymbol{\alpha}}$
    \quad
    \textbf{Dawn Lawrie}$^{\hspace{.1em}\color{RoyalBlue}\boldsymbol{\iota}}$
    \vspace{.2em}\\
    \textbf{Benjamin Van Durme}$^{\hspace{.1em}\color{RoyalBlue}\boldsymbol{\iota}}$
    \quad
    \textbf{Arman Cohan}$^{\hspace{.1em}\color{RoyalBlue}\boldsymbol{\gamma\hspace{.1em}\alpha}}$
    \quad
    \textbf{Luca Soldaini}$^{\hspace{.1em}\color{RoyalBlue}\boldsymbol{\alpha}}$
    \vspace{.5em}\\
    $^{\color{RoyalBlue}\iota\hspace{.1em}}$Johns Hopkins University
    \quad
    $^{\color{RoyalBlue}\alpha\hspace{.1em}}$Allen Institute for AI
    \quad
    $^{\color{RoyalBlue}\gamma\hspace{.1em}}$Yale University
    \vspace{.5em}\\
    \texttt{oweller@cs.jhu.edu}
    \quad
    \texttt{\{kylel, lucas\}@allenai.org}\\}

\begin{document}
\maketitle
\begin{abstract}
Using large language models (LMs) for query or document expansion can improve generalization in information retrieval.
However, it is unknown whether these techniques are universally beneficial or only effective in specific settings, such as for particular retrieval models, dataset domains, or query types.
To answer this, we conduct the first comprehensive analysis of LM-based expansion. 
We find that there exists a strong negative correlation between retriever performance and gains from expansion: expansion improves scores for weaker models, but generally harms stronger models.
We show this trend holds across a set of eleven expansion techniques, twelve datasets with diverse distribution shifts, and twenty-four retrieval models.
Through qualitative error analysis, we hypothesize that although expansions provide extra information (potentially improving recall), they add additional noise that makes it difficult to discern between the top relevant documents (thus introducing false positives).
Our results suggest the following recipe: use expansions for weaker models or when the target dataset significantly differs from training corpus in format; 
otherwise, avoid expansions to keep the relevance signal clear.\footnote{Code and data are available at \url{https://github.com/orionw/LM-expansions}}
\end{abstract}

\section{Introduction}
\extrafootertext{{\color{BrickRed}$^\ast$} Work performed during internship at AI2.}

Neural information retrieval (IR) systems routinely achieve state-of-the-art performance on tasks where labeled data is abundant~\cite{karpukhin2020dense,lin2022pretrained}. 
When limited or no data is available, neural models fine-tuned on data-rich domains are used in zero-shot manner~\cite{thakur2021beir,Rosa2022NoPL}.
However, shifts in distribution of queries and documents can negatively impact their performance~\cite{simon2023MsShift}.

\begin{figure}[t]
    \centering
    \includegraphics[width=0.9999\columnwidth,trim=0.5cm 0.5cm 0.0cm 0cm]{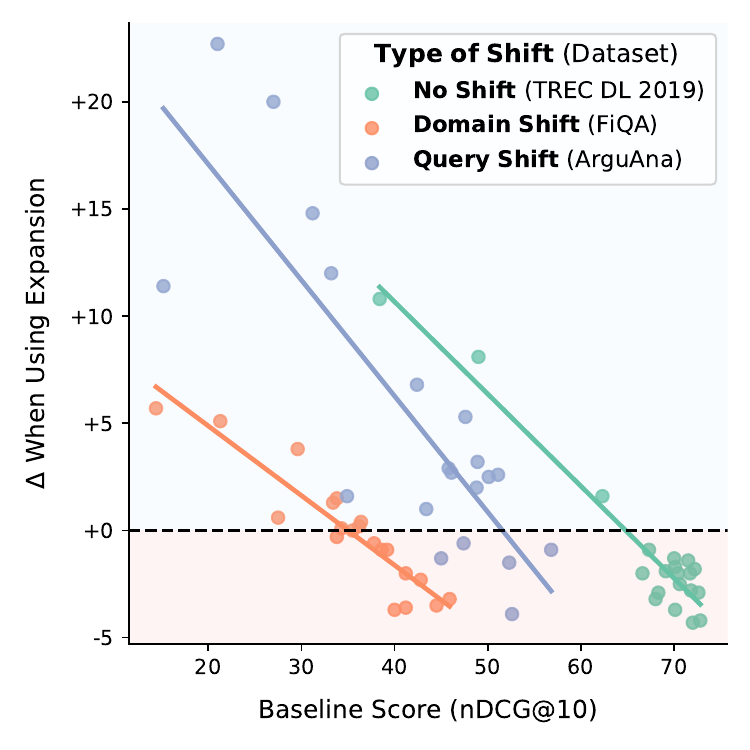}

    \caption{LM-based query and document expansion methods typically improve performance when used with weaker models, but not for stronger models. More accurate models generally lose relevance signal when expansions are provided. Each point is a value in \autoref{tab:models}.\vspace{-1em}}
    \label{fig:teaser}
\end{figure}

To mitigate this effect, language models (LMs) can be used to \textit{expand} queries or documents from unseen domains~\cite{Dai2022PromptagatorFD,gao2022precise,Jagerman2023QueryEB,jeronymo2023inpars,wang2023query2doc}. These techniques input queries and/or documents into an LM to generate additional content, which is combined with original text to facilitate relevance matching.
For example, Doc2Query \cite{nogueira2019document} uses an LM to generate likely queries for documents in the collection.
Meanwhile, HyDE~\cite{gao2022precise} uses an LM to generate a fictitious relevant document for a user query.
As LMs are often trained on more domains than typical rankers, LM-based expansion leverages this encoded knowledge to bridge out-of-distribution gaps.

IR researchers have long proposed methods to expand queries and documents \cite{rocchio1971relevance,10.1145/383952.383972,abdul2004umass}.
However, we note that LM-based expansions are qualitatively different from traditional expansion techniques. 
While the latter are largely non-parametric, using thesauri or relevance signals from the collection,\footnote{For example, pseudo relevance feedback (PRF) uses top-$k$ retrieved documents to expand queries. Thus, PRF relies on the quality of the initial retrieved set; generally, the better the retrieval, the better the expansion. We note that this is not necessarily the case for LM-based expansions/PRF: parametric knowledge encoded in model weights affect terms selected for expansion (in contrast to classic PRF that typically selects new terms from the top relevant documents from the collection).}
LM-based expansions can leverage knowledge encoded in their model weights.
Finally, while many comparative analyses of statistical expansion techniques exist \cite{hust2003towards,bhogal2007review,10.1145/2071389.2071390}, no equivalent work has been conducted for LM-based approaches.

Many works have proposed specific LM-based expansions, but these approaches are generally tested only a small subset of retrieval methods (small bi-encoder models or BM25) or only work on specific domains \cite{gao2022precise,wang2023query2doc,Zhu2023LargeLM}.
We thus seek to answer the following:
\newcommand{\rqone}{How do different models impact query and document expansion}
\newcommand{\rqtwo}{How do different distribution shifts impact these results}
\newcommand{\rqthree}{Why do expansions hurt stronger IR models}

\textbf{RQ1: \rqone\ (\autoref{sec:rq1})?}
Across all types of IR models and architectures, performance is negatively correlated with gains from expansion: after a certain score threshold these expansions generally hurt performance (as they blur the relevance signal from the original documents). 
\hspace{0.5em}

\textbf{RQ2: \rqtwo\ (\autoref{sec:rq2})?}
Our main results hold for all types of shift  -- better models are harmed by expansion -- except for long query shift, where expansions generally help most-to-all models.
\hspace{0.5em}

\textbf{RQ3: \rqthree\ (\autoref{sec:rq3})?}
We find that query and document expansions introduce new terms, potentially weakening the relevance signal of the original text.
\hspace{0.5em}

Overall, this work aims at answering the following question:
\textbf{when should one use LM-based expansions?} 
Through our investigation, we provide evidence to help practitioners answer this question. 
Our results run counter to the common intuition that query and document expansion are helpful techniques in all cases;
instead, they show that LM expansions generally \textbf{benefit weaker rankers}, but \textbf{hurt more accurate rankers}. 
Further, analysis over twelve datasets shows that whether a given model benefits from expansion varies depending on task;
datasets with pronounced distribution shifts (\textit{e.g.}, very long queries) are more likely to benefit.

\section{Experimental Settings}
\label{sec:settings}

We provide an overview of document and query expansion methods used in the reminder of the manuscript, and describe our experimental setup.

We choose expansion techniques according to two criteria: (\textit{i}) their overall performance, as claimed in papers introducing them, and (\textit{ii}) whether they can used with any retrieval model. 
While there exists more specific techniques for particular architectures, such as ColBERT-PRF \cite{wang2023colbert,wang2023effective},
we use text-based expansions from LMs to ensure generalizability of our findings.

We generate expansions using~\chatgptShort\footnote{We use version \chatgpt. 
To show that our results generalize beyond this specific language model, we include results using other open/API LMs (\gptfour, Claude V2, Llama2 70b Chat) in \autoref{app:LMs} that show the same conclusion. Prompts and example input/output can be found in Appendix~\ref{app:examples} and \ref{app:prompts}. We also explore the placement of these augmentations (should we prepend/append/replace the original query and documents?) in \autoref{app:placement} and show that this also makes little difference.} as it is inexpensive  and shows strong performance in previous work \cite{wang2023query2doc,Jagerman2023QueryEB}. Since using LMs to generate expansions for large collections would be prohibitive, we restrict our expansions to the reranking setting, e.g. the top 100 documents per query found from BM25 following \citet{asai2022task}.\footnote{As of September 2023, even just a single document expansion method using \chatgptShort on the DL Track 2019 collection would cost thousands of dollars. Thus we rerank the top 100 docs for each dataset. We show in Appendix~\ref{app:rerank_size} and Table~\ref{tab:rerank_size} that our observations hold up to 10,000 documents.} Following established practices, we use these expansions for zero-shot out-of-domain retrieval. Although it is possible that training with expansions may further increase their effectiveness, this limits their generalizability since it requires re-training retrieval models for each expansion technique and LM. 

\begin{figure*}[t]
    \centering
    \includegraphics[width=0.999\columnwidth,trim=0.5cm 0.5cm 0.0cm 0cm]{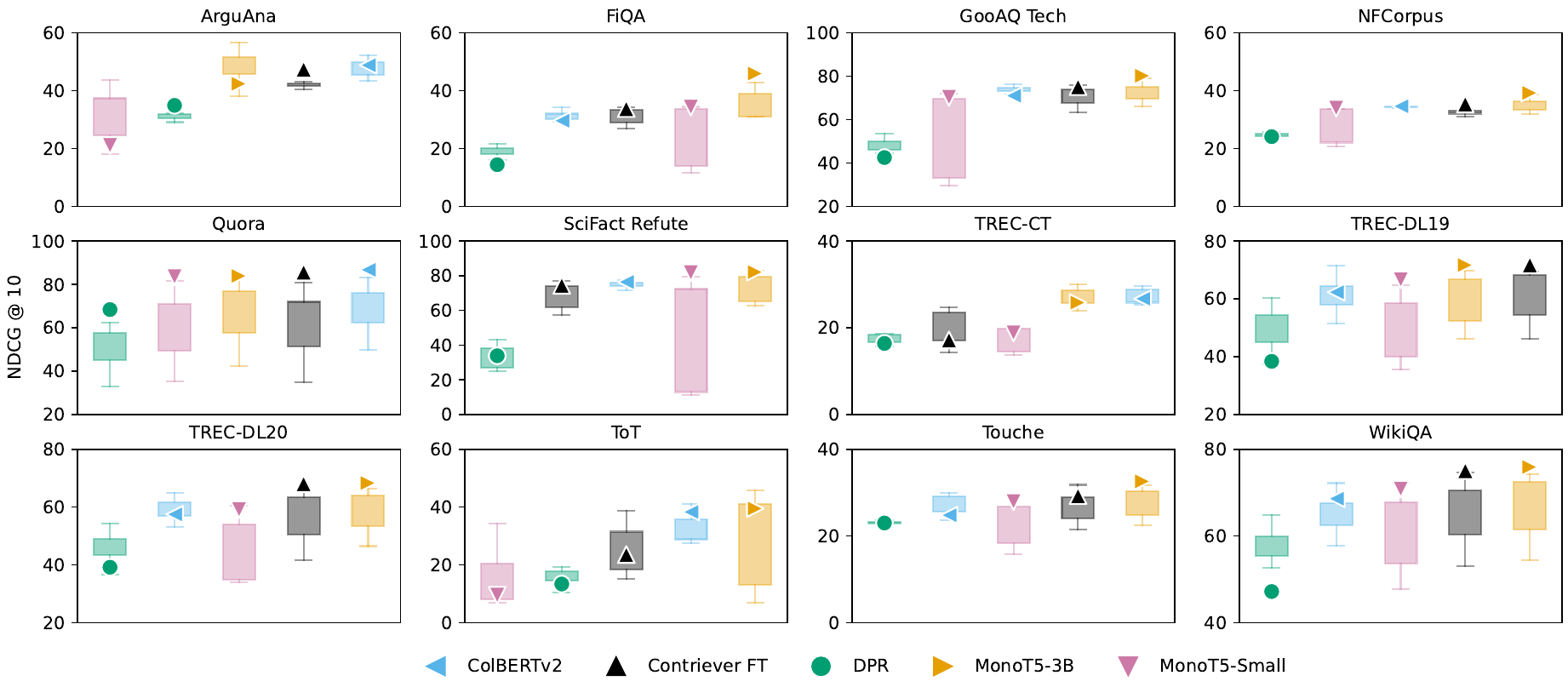}
    \caption{
    Effect of expansion over twelve datasets. 
    For each dataset, markers show base performance for models, while the boxplot indicates the range of changes in scores for document and/or query expansion.
    Across all datasets and models, we note a consistent trend:
    models with \textbf{lower base performance benefit} from expansion; 
    \textbf{higher performing rankers} generally \textbf{suffer} when expansion techniques are used. 
    \vspace{-0.5em}}
    \label{fig:models}
\end{figure*}

\begin{table*}[t!]
\centering
\small
\resizebox{\textwidth}{!}{\begin{tabular}{ll|cccc|cccc|cccc}
\toprule
\multicolumn{2}{l}{} & \multicolumn{4}{c|}{\textbf{
DL Track 2019}} & \multicolumn{4}{c}{\textbf{FiQA}} & \multicolumn{4}{c}{\textbf{Arguana}} \\
\scriptsize \textbf{Type} & \scriptsize\textbf{Model} & \scriptsize \textit{No Exp} & \scriptsize QE & \scriptsize DE & \scriptsize Both & \scriptsize \textit{No Exp} & \scriptsize QE & \scriptsize DE & \scriptsize Both & \scriptsize \textit{No Exp} & \scriptsize QE & \scriptsize DE & \scriptsize Both \\
\midrule
\parbox[t]{2mm}{\multirow{8}{*}{\rotatebox[origin=c]{90}{
 \parbox[c]{1cm}{\centering \scriptsize First Stage }}}} 
 & DPR & \gradientcell{38.4}{38.3}{73.7}{low}{high}{\opacity}{0} & \midpointgradientcell{6.6}{-6.1}{6.7}{0}{neg}{pos}{\opacity}{0} & \midpointgradientcell{3.1}{-5.5}{4.1}{0}{neg}{pos}{\opacity}{0} & \midpointgradientcell{10.8}{-9.4}{10.9}{0}{neg}{pos}{\opacity}{0} & \gradientcell{14.4}{14.3}{46.0}{low}{high}{\opacity}{0} & \midpointgradientcell{4.7}{-4.6}{4.8}{0}{neg}{pos}{\opacity}{0} & \midpointgradientcell{1.7}{-7.5}{1.8}{0}{neg}{pos}{\opacity}{0} & \midpointgradientcell{5.7}{-15.9}{5.8}{0}{neg}{pos}{\opacity}{0} & \gradientcell{34.9}{15.1}{56.9}{low}{high}{\opacity}{0} & \midpointgradientcell{-7.1}{-8.9}{22.8}{0}{neg}{pos}{\opacity}{0} & \midpointgradientcell{1.6}{-8.3}{10.9}{0}{neg}{pos}{\opacity}{0} & \midpointgradientcell{-4.4}{-9.9}{22.3}{0}{neg}{pos}{\opacity}{0} \\
 & Contriever & \gradientcell{49.0}{38.3}{73.7}{low}{high}{\opacity}{0} & \midpointgradientcell{3.5}{-6.1}{6.7}{0}{neg}{pos}{\opacity}{0} & \midpointgradientcell{4.0}{-5.5}{4.1}{0}{neg}{pos}{\opacity}{0} & \midpointgradientcell{8.1}{-9.4}{10.9}{0}{neg}{pos}{\opacity}{0} & \gradientcell{21.3}{14.3}{46.0}{low}{high}{\opacity}{0} & \midpointgradientcell{3.6}{-4.6}{4.8}{0}{neg}{pos}{\opacity}{0} & \midpointgradientcell{1.6}{-7.5}{1.8}{0}{neg}{pos}{\opacity}{0} & \midpointgradientcell{5.1}{-15.9}{5.8}{0}{neg}{pos}{\opacity}{0} & \gradientcell{45.8}{15.1}{56.9}{low}{high}{\opacity}{0} & \midpointgradientcell{-0.1}{-8.9}{22.8}{0}{neg}{pos}{\opacity}{0} & \midpointgradientcell{2.9}{-8.3}{10.9}{0}{neg}{pos}{\opacity}{0} & \midpointgradientcell{-3.2}{-9.9}{22.3}{0}{neg}{pos}{\opacity}{0} \\
& BM25 & \gradientcell{51.2}{38.3}{73.7}{low}{high}{\opacity}{0} & \midpointgradientcell{-4.0}{-6.1}{6.7}{0}{neg}{pos}{\opacity}{0} & - & - & \gradientcell{23.6}{14.3}{46.0}{low}{high}{\opacity}{0} & \midpointgradientcell{4.5}{-4.6}{4.8}{0}{neg}{pos}{\opacity}{0} & - & - & \gradientcell{30.0}{15.1}{56.9}{low}{high}{\opacity}{0} & \midpointgradientcell{-5.4}{-8.9}{22.8}{0}{neg}{pos}{\opacity}{0} & - & - \\
 & Contriever FT & \gradientcell{62.3}{38.3}{73.7}{low}{high}{\opacity}{0} & \midpointgradientcell{1.6}{-6.1}{6.7}{0}{neg}{pos}{\opacity}{0} & \midpointgradientcell{-0.2}{-5.5}{4.1}{0}{neg}{pos}{\opacity}{0} & \midpointgradientcell{0.6}{-9.4}{10.9}{0}{neg}{pos}{\opacity}{0} & \gradientcell{29.6}{14.3}{46.0}{low}{high}{\opacity}{0} & \midpointgradientcell{3.2}{-4.6}{4.8}{0}{neg}{pos}{\opacity}{0} & \midpointgradientcell{0.6}{-7.5}{1.8}{0}{neg}{pos}{\opacity}{0} & \midpointgradientcell{3.8}{-15.9}{5.8}{0}{neg}{pos}{\opacity}{0} & \gradientcell{48.8}{15.1}{56.9}{low}{high}{\opacity}{0} & \midpointgradientcell{-3.6}{-8.9}{22.8}{0}{neg}{pos}{\opacity}{0} & \midpointgradientcell{2.0}{-8.3}{10.9}{0}{neg}{pos}{\opacity}{0} & \midpointgradientcell{-2.5}{-9.9}{22.3}{0}{neg}{pos}{\opacity}{0} \\
 & E5 Base v2 & \gradientcell{67.3}{38.3}{73.7}{low}{high}{\opacity}{0} & \midpointgradientcell{-3.4}{-6.1}{6.7}{0}{neg}{pos}{\opacity}{0} & \midpointgradientcell{-0.9}{-5.5}{4.1}{0}{neg}{pos}{\opacity}{0} & \midpointgradientcell{-3.7}{-9.4}{10.9}{0}{neg}{pos}{\opacity}{0} & \gradientcell{37.8}{14.3}{46.0}{low}{high}{\opacity}{0} & \midpointgradientcell{-0.6}{-4.6}{4.8}{0}{neg}{pos}{\opacity}{0} & \midpointgradientcell{-3.8}{-7.5}{1.8}{0}{neg}{pos}{\opacity}{0} & \midpointgradientcell{-2.5}{-15.9}{5.8}{0}{neg}{pos}{\opacity}{0} & \gradientcell{51.1}{15.1}{56.9}{low}{high}{\opacity}{0} & \midpointgradientcell{-8.4}{-8.9}{22.8}{0}{neg}{pos}{\opacity}{0} & \midpointgradientcell{2.6}{-8.3}{10.9}{0}{neg}{pos}{\opacity}{0} & \midpointgradientcell{-5.7}{-9.9}{22.3}{0}{neg}{pos}{\opacity}{0} \\
 & MPNet Base v2 & \gradientcell{68.3}{38.3}{73.7}{low}{high}{\opacity}{0} & \midpointgradientcell{-6.0}{-6.1}{6.7}{0}{neg}{pos}{\opacity}{0} & \midpointgradientcell{-2.9}{-5.5}{4.1}{0}{neg}{pos}{\opacity}{0} & \midpointgradientcell{-6.8}{-9.4}{10.9}{0}{neg}{pos}{\opacity}{0} & \gradientcell{44.5}{14.3}{46.0}{low}{high}{\opacity}{0} & \midpointgradientcell{-4.1}{-4.6}{4.8}{0}{neg}{pos}{\opacity}{0} & \midpointgradientcell{-3.5}{-7.5}{1.8}{0}{neg}{pos}{\opacity}{0} & \midpointgradientcell{-5.7}{-15.9}{5.8}{0}{neg}{pos}{\opacity}{0} & \gradientcell{47.6}{15.1}{56.9}{low}{high}{\opacity}{0} & \midpointgradientcell{-5.1}{-8.9}{22.8}{0}{neg}{pos}{\opacity}{0} & \midpointgradientcell{5.3}{-8.3}{10.9}{0}{neg}{pos}{\opacity}{0} & \midpointgradientcell{-0.7}{-9.9}{22.3}{0}{neg}{pos}{\opacity}{0} \\
 & E5 Small v2 & \gradientcell{69.1}{38.3}{73.7}{low}{high}{\opacity}{0} & \midpointgradientcell{-4.8}{-6.1}{6.7}{0}{neg}{pos}{\opacity}{0} & \midpointgradientcell{-1.9}{-5.5}{4.1}{0}{neg}{pos}{\opacity}{0} & \midpointgradientcell{-6.8}{-9.4}{10.9}{0}{neg}{pos}{\opacity}{0} & \gradientcell{36.4}{14.3}{46.0}{low}{high}{\opacity}{0} & \midpointgradientcell{0.4}{-4.6}{4.8}{0}{neg}{pos}{\opacity}{0} & \midpointgradientcell{-2.9}{-7.5}{1.8}{0}{neg}{pos}{\opacity}{0} & \midpointgradientcell{-0.6}{-15.9}{5.8}{0}{neg}{pos}{\opacity}{0} & \gradientcell{46.1}{15.1}{56.9}{low}{high}{\opacity}{0} & \midpointgradientcell{-8.7}{-8.9}{22.8}{0}{neg}{pos}{\opacity}{0} & \midpointgradientcell{2.7}{-8.3}{10.9}{0}{neg}{pos}{\opacity}{0} & \midpointgradientcell{-9.8}{-9.9}{22.3}{0}{neg}{pos}{\opacity}{0} \\
 & GTE Large & \gradientcell{70.0}{38.3}{73.7}{low}{high}{\opacity}{0} & \midpointgradientcell{-4.5}{-6.1}{6.7}{0}{neg}{pos}{\opacity}{0} & \midpointgradientcell{-1.3}{-5.5}{4.1}{0}{neg}{pos}{\opacity}{0} & \midpointgradientcell{-4.5}{-9.4}{10.9}{0}{neg}{pos}{\opacity}{0} & \gradientcell{41.2}{14.3}{46.0}{low}{high}{\opacity}{0} & \midpointgradientcell{-2.0}{-4.6}{4.8}{0}{neg}{pos}{\opacity}{0} & \midpointgradientcell{-4.1}{-7.5}{1.8}{0}{neg}{pos}{\opacity}{0} & \midpointgradientcell{-3.2}{-15.9}{5.8}{0}{neg}{pos}{\opacity}{0} & \gradientcell{56.8}{15.1}{56.9}{low}{high}{\opacity}{0} & \midpointgradientcell{-8.8}{-8.9}{22.8}{0}{neg}{pos}{\opacity}{0} & \midpointgradientcell{-0.9}{-8.3}{10.9}{0}{neg}{pos}{\opacity}{0} & \midpointgradientcell{-9.0}{-9.9}{22.3}{0}{neg}{pos}{\opacity}{0} \\

 & E5 Large v2 & \gradientcell{70.1}{38.3}{73.7}{low}{high}{\opacity}{0} & \midpointgradientcell{-5.7}{-6.1}{6.7}{0}{neg}{pos}{\opacity}{0} & \midpointgradientcell{-1.7}{-5.5}{4.1}{0}{neg}{pos}{\opacity}{0} & \midpointgradientcell{-7.6}{-9.4}{10.9}{0}{neg}{pos}{\opacity}{0} & \gradientcell{38.6}{14.3}{46.0}{low}{high}{\opacity}{0} & \midpointgradientcell{-0.9}{-4.6}{4.8}{0}{neg}{pos}{\opacity}{0} & \midpointgradientcell{-2.7}{-7.5}{1.8}{0}{neg}{pos}{\opacity}{0} & \midpointgradientcell{-3.2}{-15.9}{5.8}{0}{neg}{pos}{\opacity}{0} & \gradientcell{48.9}{15.1}{56.9}{low}{high}{\opacity}{0} & \midpointgradientcell{-5.9}{-8.9}{22.8}{0}{neg}{pos}{\opacity}{0} & \midpointgradientcell{3.2}{-8.3}{10.9}{0}{neg}{pos}{\opacity}{0} & \midpointgradientcell{-3.4}{-9.9}{22.3}{0}{neg}{pos}{\opacity}{0} \\
 \midrule
\parbox[t]{2mm}{\multirow{8}{*}{\rotatebox[origin=c]{90}{
 \parbox[c]{1cm}{\centering \scriptsize Rerankers }}}} 
 & MonoT5-Small & \gradientcell{66.6}{38.3}{73.7}{low}{high}{\opacity}{0} & \midpointgradientcell{-2.0}{-6.1}{6.7}{0}{neg}{pos}{\opacity}{0} & \midpointgradientcell{-2.8}{-5.5}{4.1}{0}{neg}{pos}{\opacity}{0} & \midpointgradientcell{-2.8}{-9.4}{10.9}{0}{neg}{pos}{\opacity}{0} & \gradientcell{34.3}{14.3}{46.0}{low}{high}{\opacity}{0} & \midpointgradientcell{0.1}{-4.6}{4.8}{0}{neg}{pos}{\opacity}{0} & \midpointgradientcell{-0.6}{-7.5}{1.8}{0}{neg}{pos}{\opacity}{0} & \midpointgradientcell{-0.3}{-15.9}{5.8}{0}{neg}{pos}{\opacity}{0} & \gradientcell{21.1}{15.1}{56.9}{low}{high}{\opacity}{0} & \midpointgradientcell{22.7}{-8.9}{22.8}{0}{neg}{pos}{\opacity}{0} & \midpointgradientcell{-3.0}{-8.3}{10.9}{0}{neg}{pos}{\opacity}{0} & \midpointgradientcell{22.2}{-9.9}{22.3}{0}{neg}{pos}{\opacity}{0} \\
 & MiniLM-2-v2 & \gradientcell{68.0}{38.3}{73.7}{low}{high}{\opacity}{0} & \midpointgradientcell{-3.2}{-6.1}{6.7}{0}{neg}{pos}{\opacity}{0} & \midpointgradientcell{-4.1}{-5.5}{4.1}{0}{neg}{pos}{\opacity}{0} & \midpointgradientcell{-5.1}{-9.4}{10.9}{0}{neg}{pos}{\opacity}{0} & \gradientcell{27.5}{14.3}{46.0}{low}{high}{\opacity}{0} & \midpointgradientcell{-2.0}{-4.6}{4.8}{0}{neg}{pos}{\opacity}{0} & \midpointgradientcell{0.6}{-7.5}{1.8}{0}{neg}{pos}{\opacity}{0} & \midpointgradientcell{-15.8}{-15.9}{5.8}{0}{neg}{pos}{\opacity}{0} & \gradientcell{15.2}{15.1}{56.9}{low}{high}{\opacity}{0} & \midpointgradientcell{11.4}{-8.9}{22.8}{0}{neg}{pos}{\opacity}{0} & \midpointgradientcell{10.8}{-8.3}{10.9}{0}{neg}{pos}{\opacity}{0} & \midpointgradientcell{11.2}{-9.9}{22.3}{0}{neg}{pos}{\opacity}{0} \\
 & SPLADEv2 & \gradientcell{70.1}{38.3}{73.7}{low}{high}{\opacity}{0} & \midpointgradientcell{-4.3}{-6.1}{6.7}{0}{neg}{pos}{\opacity}{0} & \midpointgradientcell{-3.7}{-5.5}{4.1}{0}{neg}{pos}{\opacity}{0} & \midpointgradientcell{-5.6}{-9.4}{10.9}{0}{neg}{pos}{\opacity}{0} & \gradientcell{33.4}{14.3}{46.0}{low}{high}{\opacity}{0} & \midpointgradientcell{1.3}{-4.6}{4.8}{0}{neg}{pos}{\opacity}{0} & \midpointgradientcell{-0.2}{-7.5}{1.8}{0}{neg}{pos}{\opacity}{0} & \midpointgradientcell{1.2}{-15.9}{5.8}{0}{neg}{pos}{\opacity}{0} & \gradientcell{45.0}{15.1}{56.9}{low}{high}{\opacity}{0} & \midpointgradientcell{-4.5}{-8.9}{22.8}{0}{neg}{pos}{\opacity}{0} & \midpointgradientcell{-1.3}{-8.3}{10.9}{0}{neg}{pos}{\opacity}{0} & \midpointgradientcell{-4.0}{-9.9}{22.3}{0}{neg}{pos}{\opacity}{0} \\
 & MonoBERT & \gradientcell{70.4}{38.3}{73.7}{low}{high}{\opacity}{0} & \midpointgradientcell{-4.6}{-6.1}{6.7}{0}{neg}{pos}{\opacity}{0} & \midpointgradientcell{-2.0}{-5.5}{4.1}{0}{neg}{pos}{\opacity}{0} & \midpointgradientcell{-4.8}{-9.4}{10.9}{0}{neg}{pos}{\opacity}{0} & \gradientcell{36.2}{14.3}{46.0}{low}{high}{\opacity}{0} & \midpointgradientcell{0.2}{-4.6}{4.8}{0}{neg}{pos}{\opacity}{0} & \midpointgradientcell{-0.7}{-7.5}{1.8}{0}{neg}{pos}{\opacity}{0} & \midpointgradientcell{0.0}{-15.9}{5.8}{0}{neg}{pos}{\opacity}{0} & \gradientcell{50.1}{15.1}{56.9}{low}{high}{\opacity}{0} & \midpointgradientcell{-5.7}{-8.9}{22.8}{0}{neg}{pos}{\opacity}{0} & \midpointgradientcell{2.5}{-8.3}{10.9}{0}{neg}{pos}{\opacity}{0} & \midpointgradientcell{-9.3}{-9.9}{22.3}{0}{neg}{pos}{\opacity}{0} \\
 & MiniLM-4-v2 & \gradientcell{70.6}{38.3}{73.7}{low}{high}{\opacity}{0} & \midpointgradientcell{-3.0}{-6.1}{6.7}{0}{neg}{pos}{\opacity}{0} & \midpointgradientcell{-2.5}{-5.5}{4.1}{0}{neg}{pos}{\opacity}{0} & \midpointgradientcell{-4.9}{-9.4}{10.9}{0}{neg}{pos}{\opacity}{0} & \gradientcell{33.8}{14.3}{46.0}{low}{high}{\opacity}{0} & \midpointgradientcell{1.5}{-4.6}{4.8}{0}{neg}{pos}{\opacity}{0} & \midpointgradientcell{-0.3}{-7.5}{1.8}{0}{neg}{pos}{\opacity}{0} & \midpointgradientcell{1.2}{-15.9}{5.8}{0}{neg}{pos}{\opacity}{0} & \gradientcell{43.4}{15.1}{56.9}{low}{high}{\opacity}{0} & \midpointgradientcell{0.4}{-8.9}{22.8}{0}{neg}{pos}{\opacity}{0} & \midpointgradientcell{1.0}{-8.3}{10.9}{0}{neg}{pos}{\opacity}{0} & \midpointgradientcell{-0.8}{-9.9}{22.3}{0}{neg}{pos}{\opacity}{0} \\
 & MonoT5-Base & \gradientcell{71.5}{38.3}{73.7}{low}{high}{\opacity}{0} & \midpointgradientcell{-3.2}{-6.1}{6.7}{0}{neg}{pos}{\opacity}{0} & \midpointgradientcell{-1.4}{-5.5}{4.1}{0}{neg}{pos}{\opacity}{0} & \midpointgradientcell{-5.2}{-9.4}{10.9}{0}{neg}{pos}{\opacity}{0} & \gradientcell{39.2}{14.3}{46.0}{low}{high}{\opacity}{0} & \midpointgradientcell{-1.2}{-4.6}{4.8}{0}{neg}{pos}{\opacity}{0} & \midpointgradientcell{-1.2}{-7.5}{1.8}{0}{neg}{pos}{\opacity}{0} & \midpointgradientcell{-0.9}{-15.9}{5.8}{0}{neg}{pos}{\opacity}{0} & \gradientcell{27.0}{15.1}{56.9}{low}{high}{\opacity}{0} & \midpointgradientcell{20.0}{-8.9}{22.8}{0}{neg}{pos}{\opacity}{0} & \midpointgradientcell{0.7}{-8.3}{10.9}{0}{neg}{pos}{\opacity}{0} & \midpointgradientcell{18.7}{-9.9}{22.3}{0}{neg}{pos}{\opacity}{0} \\
 & MonoT5-3B & \gradientcell{71.7}{38.3}{73.7}{low}{high}{\opacity}{0} & \midpointgradientcell{-2.8}{-6.1}{6.7}{0}{neg}{pos}{\opacity}{0} & \midpointgradientcell{-2.0}{-5.5}{4.1}{0}{neg}{pos}{\opacity}{0} & \midpointgradientcell{-5.0}{-9.4}{10.9}{0}{neg}{pos}{\opacity}{0} & \gradientcell{45.9}{14.3}{46.0}{low}{high}{\opacity}{0} & \midpointgradientcell{-3.8}{-4.6}{4.8}{0}{neg}{pos}{\opacity}{0} & \midpointgradientcell{-3.2}{-7.5}{1.8}{0}{neg}{pos}{\opacity}{0} & \midpointgradientcell{-5.6}{-15.9}{5.8}{0}{neg}{pos}{\opacity}{0} & \gradientcell{42.4}{15.1}{56.9}{low}{high}{\opacity}{0} & \midpointgradientcell{6.8}{-8.9}{22.8}{0}{neg}{pos}{\opacity}{0} & \midpointgradientcell{-1.9}{-8.3}{10.9}{0}{neg}{pos}{\opacity}{0} & \midpointgradientcell{5.2}{-9.9}{22.3}{0}{neg}{pos}{\opacity}{0} \\
 & ColBERTv2 & \gradientcell{71.8}{38.3}{73.7}{low}{high}{\opacity}{0} & \midpointgradientcell{-4.2}{-6.1}{6.7}{0}{neg}{pos}{\opacity}{0} & \midpointgradientcell{-2.8}{-5.5}{4.1}{0}{neg}{pos}{\opacity}{0} & \midpointgradientcell{-6.4}{-9.4}{10.9}{0}{neg}{pos}{\opacity}{0} & \gradientcell{33.8}{14.3}{46.0}{low}{high}{\opacity}{0} & \midpointgradientcell{-0.4}{-4.6}{4.8}{0}{neg}{pos}{\opacity}{0} & \midpointgradientcell{-0.3}{-7.5}{1.8}{0}{neg}{pos}{\opacity}{0} & \midpointgradientcell{-0.7}{-15.9}{5.8}{0}{neg}{pos}{\opacity}{0} & \gradientcell{47.4}{15.1}{56.9}{low}{high}{\opacity}{0} & \midpointgradientcell{-5.2}{-8.9}{22.8}{0}{neg}{pos}{\opacity}{0} & \midpointgradientcell{-0.6}{-8.3}{10.9}{0}{neg}{pos}{\opacity}{0} & \midpointgradientcell{-4.8}{-9.9}{22.3}{0}{neg}{pos}{\opacity}{0} \\
 & MiniLM-12-v2 & \gradientcell{72.0}{38.3}{73.7}{low}{high}{\opacity}{0} & \midpointgradientcell{-4.3}{-6.1}{6.7}{0}{neg}{pos}{\opacity}{0} & \midpointgradientcell{-4.5}{-5.5}{4.1}{0}{neg}{pos}{\opacity}{0} & \midpointgradientcell{-5.6}{-9.4}{10.9}{0}{neg}{pos}{\opacity}{0} & \gradientcell{35.5}{14.3}{46.0}{low}{high}{\opacity}{0} & \midpointgradientcell{-0.4}{-4.6}{4.8}{0}{neg}{pos}{\opacity}{0} & \midpointgradientcell{-0.5}{-7.5}{1.8}{0}{neg}{pos}{\opacity}{0} & \midpointgradientcell{0.0}{-15.9}{5.8}{0}{neg}{pos}{\opacity}{0} & \gradientcell{33.2}{15.1}{56.9}{low}{high}{\opacity}{0} & \midpointgradientcell{12.0}{-8.9}{22.8}{0}{neg}{pos}{\opacity}{0} & \midpointgradientcell{1.1}{-8.3}{10.9}{0}{neg}{pos}{\opacity}{0} & \midpointgradientcell{9.8}{-9.9}{22.3}{0}{neg}{pos}{\opacity}{0} \\
 & MonoT5-Large & \gradientcell{72.2}{38.3}{73.7}{low}{high}{\opacity}{0} & \midpointgradientcell{-4.0}{-6.1}{6.7}{0}{neg}{pos}{\opacity}{0} & \midpointgradientcell{-1.8}{-5.5}{4.1}{0}{neg}{pos}{\opacity}{0} & \midpointgradientcell{-5.6}{-9.4}{10.9}{0}{neg}{pos}{\opacity}{0} & \gradientcell{42.8}{14.3}{46.0}{low}{high}{\opacity}{0} & \midpointgradientcell{-2.3}{-4.6}{4.8}{0}{neg}{pos}{\opacity}{0} & \midpointgradientcell{-2.3}{-7.5}{1.8}{0}{neg}{pos}{\opacity}{0} & \midpointgradientcell{-3.1}{-15.9}{5.8}{0}{neg}{pos}{\opacity}{0} & \gradientcell{31.2}{15.1}{56.9}{low}{high}{\opacity}{0} & \midpointgradientcell{14.8}{-8.9}{22.8}{0}{neg}{pos}{\opacity}{0} & \midpointgradientcell{-2.0}{-8.3}{10.9}{0}{neg}{pos}{\opacity}{0} & \midpointgradientcell{14.8}{-9.9}{22.3}{0}{neg}{pos}{\opacity}{0} \\
 & LLAMA & \gradientcell{72.6}{38.3}{73.7}{low}{high}{\opacity}{0} & \midpointgradientcell{-2.9}{-6.1}{6.7}{0}{neg}{pos}{\opacity}{0} & \midpointgradientcell{-4.9}{-5.5}{4.1}{0}{neg}{pos}{\opacity}{0} & \midpointgradientcell{-7.7}{-9.4}{10.9}{0}{neg}{pos}{\opacity}{0} & \gradientcell{40.0}{14.3}{46.0}{low}{high}{\opacity}{0} & \midpointgradientcell{-3.7}{-4.6}{4.8}{0}{neg}{pos}{\opacity}{0} & \midpointgradientcell{-4.9}{-7.5}{1.8}{0}{neg}{pos}{\opacity}{0} & \midpointgradientcell{-5.8}{-15.9}{5.8}{0}{neg}{pos}{\opacity}{0} & \gradientcell{52.6}{15.1}{56.9}{low}{high}{\opacity}{0} & \midpointgradientcell{-3.9}{-8.9}{22.8}{0}{neg}{pos}{\opacity}{0} & \midpointgradientcell{-6.9}{-8.3}{10.9}{0}{neg}{pos}{\opacity}{0} & \midpointgradientcell{-9.4}{-9.9}{22.3}{0}{neg}{pos}{\opacity}{0} \\
 & LLAMAv2 & \gradientcell{72.8}{38.3}{73.7}{low}{high}{\opacity}{0} & \midpointgradientcell{-4.2}{-6.1}{6.7}{0}{neg}{pos}{\opacity}{0} & \midpointgradientcell{-4.9}{-5.5}{4.1}{0}{neg}{pos}{\opacity}{0} & \midpointgradientcell{-9.3}{-9.4}{10.9}{0}{neg}{pos}{\opacity}{0} & \gradientcell{41.1}{14.3}{46.0}{low}{high}{\opacity}{0} & \midpointgradientcell{-3.6}{-4.6}{4.8}{0}{neg}{pos}{\opacity}{0} & \midpointgradientcell{-7.4}{-7.5}{1.8}{0}{neg}{pos}{\opacity}{0} & \midpointgradientcell{-7.9}{-15.9}{5.8}{0}{neg}{pos}{\opacity}{0} & \gradientcell{52.3}{15.1}{56.9}{low}{high}{\opacity}{0} & \midpointgradientcell{-1.5}{-8.9}{22.8}{0}{neg}{pos}{\opacity}{0} & \midpointgradientcell{-8.2}{-8.3}{10.9}{0}{neg}{pos}{\opacity}{0} & \midpointgradientcell{-7.0}{-9.9}{22.3}{0}{neg}{pos}{\opacity}{0} \\
 & LLAMAv2-13B & \gradientcell{73.6}{38.3}{73.7}{low}{high}{\opacity}{0} & \midpointgradientcell{-4.5}{-6.1}{6.7}{0}{neg}{pos}{\opacity}{0} & \midpointgradientcell{-5.4}{-5.5}{4.1}{0}{neg}{pos}{\opacity}{0} & \midpointgradientcell{-7.3}{-9.4}{10.9}{0}{neg}{pos}{\opacity}{0} & \gradientcell{41.2}{14.3}{46.0}{low}{high}{\opacity}{0} & \midpointgradientcell{-4.5}{-4.6}{4.8}{0}{neg}{pos}{\opacity}{0} & \midpointgradientcell{-4.9}{-7.5}{1.8}{0}{neg}{pos}{\opacity}{0} & \midpointgradientcell{-7.0}{-15.9}{5.8}{0}{neg}{pos}{\opacity}{0} & \gradientcell{49.4}{15.1}{56.9}{low}{high}{\opacity}{0} & \midpointgradientcell{-2.1}{-8.9}{22.8}{0}{neg}{pos}{\opacity}{0} & \midpointgradientcell{-6.0}{-8.3}{10.9}{0}{neg}{pos}{\opacity}{0} & \midpointgradientcell{-4.9}{-9.9}{22.3}{0}{neg}{pos}{\opacity}{0} \\
\bottomrule
\end{tabular}}
\caption{Best expansion strategies across different models. \textit{QE} stands for query expansion (Q-LM PRF), \textit{DE} for document expansion (Doc2Query), and \textit{Both} for the combination (Q-LM PRF + Doc2Query). Colors indicate a \hlpos{positive} or \hlneg{negative} delta over \hlbase{scores for no expansion}. Models with higher base scores are generally harmed by expansions while weaker models benefit from them. Llama models follow MonoT5 in fine-tuning on MSMarco.\vspace{-0.5em}}
\label{tab:models}
\end{table*}

\subsection{Query Expansion}
We use three types of query expansion, selecting the best methods from previous work. 

\paragraph{HyDE from \citet{gao2022precise}} provides task-specific instructions for the LM to generate a document that would answer that question. We use prompts from their work when available.

\paragraph{Chain of Thought from \citet{Jagerman2023QueryEB}} was inspired by \citet{wei2022chain}; 
it prompts the model to reason before giving the answer. 
The step-by-step reasoning is then used to expand the original query.
Many works have shown the effectiveness of this approach~\citep{Jagerman2023QueryEB,he2022rethinking,trivedi2022interleaving}.

\paragraph{LM-based Pseudo Relevance Feedback (Q-LM PRF).} PRF is a classical IR method to expand a query using terms from top retrieved documents.
We use an LM to generate a list of terms from the top 3 documents ranked by a bi-encoder model (Contriever). 
Through a second invocation, the LM updates the query to include the new terms.
LM-aided PRF has been shown to be broadly effective~\cite{Mackie2023GenerativeRF,Jagerman2023QueryEB}.

\begin{table*}[t!]
    \resizebox{\textwidth}{!}{\begin{tabular}{l|l|ccc|cc}
        \toprule
           \textbf{Axis} & \textbf{Dataset} & \textbf{\# Queries} & \textbf{\# Docs} & \textbf{Avg. Judged/Q} & \textbf{Q Len} & \textbf{D Len} \\
           \toprule

     \multirow{2}{*}{In-Domain}        & TREC DL Track 2019 \cite{craswell2020overview}     & 43 & 8,841,823 & 212.5 & 5.4 & 56.6 \\
          & TREC DL Track 2020 \cite{craswell2021overview}  & 54 & 8,841,823 & 207.9 & 6.0 & 56.6 \\
         \midrule
   
           \multirow{3}{*}{Domain Shift}     & FiQA-2018 \cite{10.1145/3184558.3192301}    & 648 & 57,600 & 2.6 & 10.9 & 137.4 \\
          & Gooaq Technical  \cite{khashabi2021gooaq}     & 1,000 & 4,086 & 1.0 & 8.3 & 44.5 \\
         & NFCorpus  \cite{boteva2016}    & 323 & 3,633 & 38.2 & 3.3 & 233.5 \\ 
         \midrule

        \multirow{2}{*}{Relevance Shift }    & Touch\'e-2020 \cite{stein:2020v}   & 49 & 382,545 & 19.0 & 6.6 & 293.7 \\
             & SciFact Refute  \cite{wadden-etal-2020-fact}   & 64 & 5,183 & 1.2 & 12.1 & 214.8 \\
\midrule

         \multirow{3}{*}{Long Query Shift} & Tip of My Tongue \cite{lin2023decomposing} & 2,272 & 1,877 & 1.0 & 144.3 & 100.5 \\
             & TREC Clinical Trials '21 \cite{roberts2021overview} & 75 & 375,580 & 348.8 & 133.3 & 919.5 \\
         & ArguAna \cite{wachsmuth:2018a} & 1,406 & 8,674 & 1.0 & 197.1 & 170.3 \\
          \midrule

            \multirow{2}{*}{Short Doc Shift}        &  WikiQA \cite{yang2015wikiqa} & 369 & 26,196 & 1.2 & 6.3 & 25.1 \\
           & Quora \cite{iyer2017quora}  & 10,000 & 522,931 & 1.6 & 9.5 & 12.5 \\
    
    \bottomrule
    \end{tabular}}
    \caption{Statistics of datasets in this work. Avg. Judged/Q is the number of relevant documents per query. Length is measured in words. The TREC DL Track uses the MS MARCO dataset~\cite{nguyen2016ms}.\label{tab:dataset_stats}}
   
\end{table*}
\begin{table*}[]
\centering
\small
\begin{tabular}{ll|ccc|ccc}
\toprule
\multicolumn{2}{l}{} & \multicolumn{3}{|c|}{\textbf{
DL 2019 Track}} & \multicolumn{3}{c}{\textbf{DL 2020 Track}} \\
\scriptsize \textbf{Type} & \scriptsize\textbf{Model}  & \scriptsize DPR & \scriptsize Contriever FT & \scriptsize MonoT5-3B & \scriptsize DPR & \scriptsize Contriever FT & \scriptsize MonoT5-3B \\
\midrule
$-$
 & \textit{No Expansion} & \gradientcell{38.4}{38.3}{71.8}{low}{high}{\opacity}{0} & \gradientcell{62.3}{38.3}{71.8}{low}{high}{\opacity}{0} & \gradientcell{71.7}{38.3}{71.8}{low}{high}{\opacity}{0} & \gradientcell{39.2}{38.3}{71.8}{low}{high}{\opacity}{0} & \gradientcell{57.5}{38.3}{71.8}{low}{high}{\opacity}{0} & \gradientcell{68.3}{38.3}{71.8}{low}{high}{\opacity}{0} \\
\midrule
\parbox[t]{2mm}{\multirow{3}{*}{\rotatebox[origin=c]{90}{
 \parbox[c]{1cm}{\centering \scriptsize Query }}}} 
 & HyDE & \midpointgradientcell{18.8}{-1.2}{22.0}{0}{neg}{pos}{\opacity}{0} & \midpointgradientcell{9.3}{-15.6}{9.4}{0}{neg}{pos}{\opacity}{1} & \midpointgradientcell{-4.0}{-25.1}{-1.1}{0}{neg}{pos}{\opacity}{0} & \midpointgradientcell{13.2}{-2.7}{15.1}{0}{neg}{pos}{\opacity}{0} & \midpointgradientcell{7.4}{-9.2}{7.5}{0}{neg}{pos}{\opacity}{1} & \midpointgradientcell{-5.8}{-21.9}{-1.8}{0}{neg}{pos}{\opacity}{0} \\
 & CoT & \midpointgradientcell{12.6}{-1.2}{22.0}{0}{neg}{pos}{\opacity}{0} & \midpointgradientcell{2.7}{-15.6}{9.4}{0}{neg}{pos}{\opacity}{0} & \midpointgradientcell{-6.7}{-25.1}{-1.1}{0}{neg}{pos}{\opacity}{0} & \midpointgradientcell{5.5}{-2.7}{15.1}{0}{neg}{pos}{\opacity}{0} & \midpointgradientcell{4.2}{-9.2}{7.5}{0}{neg}{pos}{\opacity}{0} & \midpointgradientcell{-9.3}{-21.9}{-1.8}{0}{neg}{pos}{\opacity}{0} \\
 & Q-LM PRF & \midpointgradientcell{6.6}{-1.2}{22.0}{0}{neg}{pos}{\opacity}{0} & \midpointgradientcell{1.6}{-15.6}{9.4}{0}{neg}{pos}{\opacity}{0} & \midpointgradientcell{-2.2}{-25.1}{-1.1}{0}{neg}{pos}{\opacity}{0} & \midpointgradientcell{6.3}{-2.7}{15.1}{0}{neg}{pos}{\opacity}{0} & \midpointgradientcell{2.7}{-9.2}{7.5}{0}{neg}{pos}{\opacity}{0} & \midpointgradientcell{-3.0}{-21.9}{-1.8}{0}{neg}{pos}{\opacity}{0} \\
\midrule
\parbox[t]{2mm}{\multirow{2}{*}{\rotatebox[origin=c]{90}{
 \parbox[c]{0.5cm}{\centering \scriptsize Doc }}}} 
 & D2Q & \midpointgradientcell{3.1}{-1.2}{22.0}{0}{neg}{pos}{\opacity}{0} & \midpointgradientcell{-0.2}{-15.6}{9.4}{0}{neg}{pos}{\opacity}{0} & \midpointgradientcell{-1.2}{-25.1}{-1.1}{0}{neg}{pos}{\opacity}{1} & \midpointgradientcell{3.1}{-2.7}{15.1}{0}{neg}{pos}{\opacity}{0} & \midpointgradientcell{1.3}{-9.2}{7.5}{0}{neg}{pos}{\opacity}{0} & \midpointgradientcell{-1.9}{-21.9}{-1.8}{0}{neg}{pos}{\opacity}{1} \\
 & D-LM PRF & \midpointgradientcell{-1.1}{-1.2}{22.0}{0}{neg}{pos}{\opacity}{0} & \midpointgradientcell{-15.5}{-15.6}{9.4}{0}{neg}{pos}{\opacity}{0} & \midpointgradientcell{-23.6}{-25.1}{-1.1}{0}{neg}{pos}{\opacity}{0} & \midpointgradientcell{-2.6}{-2.7}{15.1}{0}{neg}{pos}{\opacity}{0} & \midpointgradientcell{-9.1}{-9.2}{7.5}{0}{neg}{pos}{\opacity}{0} & \midpointgradientcell{-19.3}{-21.9}{-1.8}{0}{neg}{pos}{\opacity}{0} \\
\midrule
\parbox[t]{2mm}{\multirow{6}{*}{\rotatebox[origin=c]{90}{
 \parbox[c]{1cm}{\centering \scriptsize Both }}}} 
 & HyDE + D2Q & \midpointgradientcell{21.9}{-1.2}{22.0}{0}{neg}{pos}{\opacity}{1} & \midpointgradientcell{9.0}{-15.6}{9.4}{0}{neg}{pos}{\opacity}{0} & \midpointgradientcell{-4.5}{-25.1}{-1.1}{0}{neg}{pos}{\opacity}{0} & \midpointgradientcell{15.0}{-2.7}{15.1}{0}{neg}{pos}{\opacity}{1} & \midpointgradientcell{6.2}{-9.2}{7.5}{0}{neg}{pos}{\opacity}{0} & \midpointgradientcell{-5.4}{-21.9}{-1.8}{0}{neg}{pos}{\opacity}{0} \\
 & CoT + D2Q & \midpointgradientcell{15.1}{-1.2}{22.0}{0}{neg}{pos}{\opacity}{0} & \midpointgradientcell{0.8}{-15.6}{9.4}{0}{neg}{pos}{\opacity}{0} & \midpointgradientcell{-7.3}{-25.1}{-1.1}{0}{neg}{pos}{\opacity}{0} & \midpointgradientcell{7.2}{-2.7}{15.1}{0}{neg}{pos}{\opacity}{0} & \midpointgradientcell{4.2}{-9.2}{7.5}{0}{neg}{pos}{\opacity}{0} & \midpointgradientcell{-8.1}{-21.9}{-1.8}{0}{neg}{pos}{\opacity}{0} \\
 & Q-LM PRF + D2Q & \midpointgradientcell{10.8}{-1.2}{22.0}{0}{neg}{pos}{\opacity}{0} & \midpointgradientcell{0.6}{-15.6}{9.4}{0}{neg}{pos}{\opacity}{0} & \midpointgradientcell{-4.2}{-25.1}{-1.1}{0}{neg}{pos}{\opacity}{0} & \midpointgradientcell{8.1}{-2.7}{15.1}{0}{neg}{pos}{\opacity}{0} & \midpointgradientcell{3.7}{-9.2}{7.5}{0}{neg}{pos}{\opacity}{0} & \midpointgradientcell{-3.3}{-21.9}{-1.8}{0}{neg}{pos}{\opacity}{0} \\
 & HyDE + D-LM PRF & \midpointgradientcell{16.7}{-1.2}{22.0}{0}{neg}{pos}{\opacity}{0} & \midpointgradientcell{-3.1}{-15.6}{9.4}{0}{neg}{pos}{\opacity}{0} & \midpointgradientcell{-22.8}{-25.1}{-1.1}{0}{neg}{pos}{\opacity}{0} & \midpointgradientcell{11.4}{-2.7}{15.1}{0}{neg}{pos}{\opacity}{0} & \midpointgradientcell{1.2}{-9.2}{7.5}{0}{neg}{pos}{\opacity}{0} & \midpointgradientcell{-17.9}{-21.9}{-1.8}{0}{neg}{pos}{\opacity}{0} \\
 & CoT + D-LM PRF & \midpointgradientcell{10.9}{-1.2}{22.0}{0}{neg}{pos}{\opacity}{0} & \midpointgradientcell{-10.9}{-15.6}{9.4}{0}{neg}{pos}{\opacity}{0} & \midpointgradientcell{-25.0}{-25.1}{-1.1}{0}{neg}{pos}{\opacity}{0} & \midpointgradientcell{4.1}{-2.7}{15.1}{0}{neg}{pos}{\opacity}{0} & \midpointgradientcell{-4.4}{-9.2}{7.5}{0}{neg}{pos}{\opacity}{0} & \midpointgradientcell{-21.8}{-21.9}{-1.8}{0}{neg}{pos}{\opacity}{0} \\
 & Q+D LM PRF & \midpointgradientcell{6.8}{-1.2}{22.0}{0}{neg}{pos}{\opacity}{0} & \midpointgradientcell{-5.6}{-15.6}{9.4}{0}{neg}{pos}{\opacity}{0} & \midpointgradientcell{-14.4}{-25.1}{-1.1}{0}{neg}{pos}{\opacity}{0} & \midpointgradientcell{4.5}{-2.7}{15.1}{0}{neg}{pos}{\opacity}{0} & \midpointgradientcell{-2.4}{-9.2}{7.5}{0}{neg}{pos}{\opacity}{0} & \midpointgradientcell{-11.8}{-21.9}{-1.8}{0}{neg}{pos}{\opacity}{0} \\
\bottomrule
\end{tabular}
\caption{In-Domain performance on the TREC Deep Learning Tracks, according to various types of expansions, showing that expansion typically helps weaker models (like DPR) but hurts stronger models (especially large reranker models like MonoT5-3B). Colors indicate a \hlpos{positive} or \hlneg{negative} delta over \hlbase{scores for no expansion}.\vspace{-1em}}
\label{tab:msmarco}
\end{table*}

\subsection{Document Expansion}

\paragraph{Doc2Query.} There are fewer widespread LM document expansion techniques, with the main one being Doc2Query~\cite{nogueira2019document}. 
Work has found that improving the question generation model results in higher scores, hence we use ChatGPT instead of T5 for our experiments \cite{nogueira2019doc2query}. 
See \autoref{app:LMs} for results using alternative LMs for document expansion.

\paragraph{LM-based Document PRF (D-LM PRF).} Similar to the Q-LM PRF technique above, we propose a document expansion that draws pseudo-relevance from \textit{related queries} instead of related documents. In this setting, where there exists a set of unjudged user queries, we show the LM the top 5 most-similar queries and ask it to expand the original document to better answer the relevant queries.

\section{RQ1: How Do Different Models Impact Query and Document Expansion?}
\label{sec:rq1}
\paragraph{Experimental Setting}
To understand the efficacy of LM-based expansions, we employ a wide variety of neural retrieval models: DPR~\cite{karpukhin2020dense}; ColBERT v2~\cite{santhanam2021colbertv2}; SPLADE v2~\cite{formal2021splade}; MonoBERT~\cite{nogueira2019multi}; several MonoT5~\cite{nogueira2020document}, E5~\cite{wang2022text}, and MiniLM models~\cite{wang2020minilm}; GTE~\cite{li2023towards}; \texttt{all-mpnet-v2-base}~\cite{reimers-2019-sentence-bert}; Llama 1 \& 2 models~\cite{touvron2023llama,touvron2023llama2}, which we fine-tune on MS MARCO.

\setlength\tabcolsep{3 pt} %
\begin{table*}[t!]
\centering
\small
\resizebox{\textwidth}{!}{\begin{tabular}{ll|ccc|ccc|ccc}
\toprule
\multicolumn{2}{l}{} & \multicolumn{3}{|c|}{\textbf{
FiQA-2018}} & \multicolumn{3}{c}{\textbf{GooAQ Technical}} & \multicolumn{3}{|c|}{\textbf{NFCorpus}} \\
\scriptsize \textbf{Type} & \scriptsize\textbf{Model} & \scriptsize DPR  & \scriptsize Contriever FT & \scriptsize MonoT5-3B  & \scriptsize DPR  & \scriptsize Contriever FT & \scriptsize MonoT5-3B & \scriptsize DPR  & \scriptsize Contriever FT & \scriptsize MonoT5-3B \\
\midrule

    & \textit{No Expansion} & \gradientcell{14.4}{14.3}{80.3}{low}{high}{\opacity}{0} & \gradientcell{29.6}{14.3}{80.3}{low}{high}{\opacity}{0} & \gradientcell{45.9}{14.3}{80.3}{low}{high}{\opacity}{0} & \gradientcell{42.5}{14.3}{80.3}{low}{high}{\opacity}{0} & \gradientcell{71.0}{14.3}{80.3}{low}{high}{\opacity}{0} & \gradientcell{80.2}{14.3}{80.3}{low}{high}{\opacity}{0} & \gradientcell{24.1}{14.3}{80.3}{low}{high}{\opacity}{0} & \gradientcell{34.6}{14.3}{80.3}{low}{high}{\opacity}{0} & \gradientcell{39.2}{14.3}{80.3}{low}{high}{\opacity}{0} \\
\midrule
\parbox[t]{2mm}{\multirow{3}{*}{\rotatebox[origin=c]{90}{
 \parbox[c]{1cm}{\centering \scriptsize Query }}}} 
 & HyDE & \midpointgradientcell{3.6}{1.6}{7.4}{0}{neg}{pos}{\opacity}{0} & \midpointgradientcell{-0.3}{-0.4}{4.7}{0}{neg}{pos}{\opacity}{0} & \midpointgradientcell{-14.7}{-15.0}{-3.1}{0}{neg}{pos}{\opacity}{0} & \midpointgradientcell{3.1}{1.9}{11.0}{0}{neg}{pos}{\opacity}{0} & \midpointgradientcell{3.8}{0.5}{5.3}{0}{neg}{pos}{\opacity}{0} & \midpointgradientcell{-10.0}{-14.3}{-1.0}{0}{neg}{pos}{\opacity}{0} & \midpointgradientcell{0.3}{-0.8}{1.7}{0}{neg}{pos}{\opacity}{0} & \midpointgradientcell{0.0}{-0.7}{0.7}{0}{neg}{pos}{\opacity}{0} & \midpointgradientcell{-5.9}{-7.3}{-0.4}{0}{neg}{pos}{\opacity}{0} \\
 & CoT & \midpointgradientcell{3.6}{1.6}{7.4}{0}{neg}{pos}{\opacity}{0} & \midpointgradientcell{0.4}{-0.4}{4.7}{0}{neg}{pos}{\opacity}{0} & \midpointgradientcell{-13.2}{-15.0}{-3.1}{0}{neg}{pos}{\opacity}{0} & \midpointgradientcell{2.0}{1.9}{11.0}{0}{neg}{pos}{\opacity}{0} & \midpointgradientcell{2.1}{0.5}{5.3}{0}{neg}{pos}{\opacity}{0} & \midpointgradientcell{-9.7}{-14.3}{-1.0}{0}{neg}{pos}{\opacity}{0} & \midpointgradientcell{-0.7}{-0.8}{1.7}{0}{neg}{pos}{\opacity}{0} & \midpointgradientcell{-0.6}{-0.7}{0.7}{0}{neg}{pos}{\opacity}{0} & \midpointgradientcell{-4.5}{-7.3}{-0.4}{0}{neg}{pos}{\opacity}{0} \\
 & Q-LM PRF & \midpointgradientcell{4.7}{1.6}{7.4}{0}{neg}{pos}{\opacity}{0} & \midpointgradientcell{3.2}{-0.4}{4.7}{0}{neg}{pos}{\opacity}{0} & \midpointgradientcell{-3.8}{-15.0}{-3.1}{0}{neg}{pos}{\opacity}{0} & \midpointgradientcell{6.4}{1.9}{11.0}{0}{neg}{pos}{\opacity}{0} & \midpointgradientcell{1.9}{0.5}{5.3}{0}{neg}{pos}{\opacity}{0} & \midpointgradientcell{-3.4}{-14.3}{-1.0}{0}{neg}{pos}{\opacity}{0} & \midpointgradientcell{0.2}{-0.8}{1.7}{0}{neg}{pos}{\opacity}{0} & \midpointgradientcell{-0.4}{-0.7}{0.7}{0}{neg}{pos}{\opacity}{0} & \midpointgradientcell{-2.7}{-7.3}{-0.4}{0}{neg}{pos}{\opacity}{0} \\
\midrule
\parbox[t]{2mm}{\multirow{2}{*}{\rotatebox[origin=c]{90}{
 \parbox[c]{0.5cm}{\centering \scriptsize Doc }}}} 
 & D2Q & \midpointgradientcell{1.7}{1.6}{7.4}{0}{neg}{pos}{\opacity}{0} & \midpointgradientcell{0.6}{-0.4}{4.7}{0}{neg}{pos}{\opacity}{0} & \midpointgradientcell{-3.2}{-15.0}{-3.1}{0}{neg}{pos}{\opacity}{1} & \midpointgradientcell{6.4}{1.9}{11.0}{0}{neg}{pos}{\opacity}{0} & \midpointgradientcell{3.0}{0.5}{5.3}{0}{neg}{pos}{\opacity}{0} & \midpointgradientcell{-1.1}{-14.3}{-1.0}{0}{neg}{pos}{\opacity}{1} & \midpointgradientcell{1.3}{-0.8}{1.7}{0}{neg}{pos}{\opacity}{0} & \midpointgradientcell{0.6}{-0.7}{0.7}{0}{neg}{pos}{\opacity}{1} & \midpointgradientcell{-0.5}{-7.3}{-0.4}{0}{neg}{pos}{\opacity}{1} \\
 & D-LM PRF & \midpointgradientcell{3.3}{1.6}{7.4}{0}{neg}{pos}{\opacity}{0} & \midpointgradientcell{1.6}{-0.4}{4.7}{0}{neg}{pos}{\opacity}{0} & \midpointgradientcell{-12.5}{-15.0}{-3.1}{0}{neg}{pos}{\opacity}{0} & \midpointgradientcell{3.8}{1.9}{11.0}{0}{neg}{pos}{\opacity}{0} & \midpointgradientcell{0.6}{0.5}{5.3}{0}{neg}{pos}{\opacity}{0} & \midpointgradientcell{-11.4}{-14.3}{-1.0}{0}{neg}{pos}{\opacity}{0} & \midpointgradientcell{0.3}{-0.8}{1.7}{0}{neg}{pos}{\opacity}{0} & \midpointgradientcell{-0.3}{-0.7}{0.7}{0}{neg}{pos}{\opacity}{0} & \midpointgradientcell{-0.7}{-7.3}{-0.4}{0}{neg}{pos}{\opacity}{0} \\
\midrule
\parbox[t]{2mm}{\multirow{6}{*}{\rotatebox[origin=c]{90}{
 \parbox[c]{1cm}{\centering \scriptsize Both }}}} 
 & HyDE + D2Q & \midpointgradientcell{4.5}{1.6}{7.4}{0}{neg}{pos}{\opacity}{0} & \midpointgradientcell{0.4}{-0.4}{4.7}{0}{neg}{pos}{\opacity}{0} & \midpointgradientcell{-14.8}{-15.0}{-3.1}{0}{neg}{pos}{\opacity}{0} & \midpointgradientcell{8.2}{1.9}{11.0}{0}{neg}{pos}{\opacity}{0} & \midpointgradientcell{5.2}{0.5}{5.3}{0}{neg}{pos}{\opacity}{1} & \midpointgradientcell{-7.4}{-14.3}{-1.0}{0}{neg}{pos}{\opacity}{0} & \midpointgradientcell{1.6}{-0.8}{1.7}{0}{neg}{pos}{\opacity}{1} & \midpointgradientcell{0.1}{-0.7}{0.7}{0}{neg}{pos}{\opacity}{0} & \midpointgradientcell{-7.2}{-7.3}{-0.4}{0}{neg}{pos}{\opacity}{0} \\
 & CoT + D2Q & \midpointgradientcell{4.4}{1.6}{7.4}{0}{neg}{pos}{\opacity}{0} & \midpointgradientcell{0.2}{-0.4}{4.7}{0}{neg}{pos}{\opacity}{0} & \midpointgradientcell{-13.4}{-15.0}{-3.1}{0}{neg}{pos}{\opacity}{0} & \midpointgradientcell{7.2}{1.9}{11.0}{0}{neg}{pos}{\opacity}{0} & \midpointgradientcell{3.8}{0.5}{5.3}{0}{neg}{pos}{\opacity}{0} & \midpointgradientcell{-6.9}{-14.3}{-1.0}{0}{neg}{pos}{\opacity}{0} & \midpointgradientcell{0.8}{-0.8}{1.7}{0}{neg}{pos}{\opacity}{0} & \midpointgradientcell{0.0}{-0.7}{0.7}{0}{neg}{pos}{\opacity}{0} & \midpointgradientcell{-5.6}{-7.3}{-0.4}{0}{neg}{pos}{\opacity}{0} \\
 & Q-LM PRF + D2Q & \midpointgradientcell{5.7}{1.6}{7.4}{0}{neg}{pos}{\opacity}{0} & \midpointgradientcell{3.8}{-0.4}{4.7}{0}{neg}{pos}{\opacity}{0} & \midpointgradientcell{-5.6}{-15.0}{-3.1}{0}{neg}{pos}{\opacity}{0} & \midpointgradientcell{10.9}{1.9}{11.0}{0}{neg}{pos}{\opacity}{1} & \midpointgradientcell{4.2}{0.5}{5.3}{0}{neg}{pos}{\opacity}{0} & \midpointgradientcell{-4.1}{-14.3}{-1.0}{0}{neg}{pos}{\opacity}{0} & \midpointgradientcell{1.4}{-0.8}{1.7}{0}{neg}{pos}{\opacity}{0} & \midpointgradientcell{-0.1}{-0.7}{0.7}{0}{neg}{pos}{\opacity}{0} & \midpointgradientcell{-3.0}{-7.3}{-0.4}{0}{neg}{pos}{\opacity}{0} \\
 & HyDE + D-LM PRF & \midpointgradientcell{5.8}{1.6}{7.4}{0}{neg}{pos}{\opacity}{0} & \midpointgradientcell{1.2}{-0.4}{4.7}{0}{neg}{pos}{\opacity}{0} & \midpointgradientcell{-14.8}{-15.0}{-3.1}{0}{neg}{pos}{\opacity}{0} & \midpointgradientcell{5.3}{1.9}{11.0}{0}{neg}{pos}{\opacity}{0} & \midpointgradientcell{2.7}{0.5}{5.3}{0}{neg}{pos}{\opacity}{0} & \midpointgradientcell{-14.2}{-14.3}{-1.0}{0}{neg}{pos}{\opacity}{0} & \midpointgradientcell{0.8}{-0.8}{1.7}{0}{neg}{pos}{\opacity}{0} & \midpointgradientcell{0.1}{-0.7}{0.7}{0}{neg}{pos}{\opacity}{0} & \midpointgradientcell{-6.3}{-7.3}{-0.4}{0}{neg}{pos}{\opacity}{0} \\
 & CoT + D-LM PRF & \midpointgradientcell{6.2}{1.6}{7.4}{0}{neg}{pos}{\opacity}{0} & \midpointgradientcell{1.7}{-0.4}{4.7}{0}{neg}{pos}{\opacity}{0} & \midpointgradientcell{-14.9}{-15.0}{-3.1}{0}{neg}{pos}{\opacity}{0} & \midpointgradientcell{3.6}{1.9}{11.0}{0}{neg}{pos}{\opacity}{0} & \midpointgradientcell{1.9}{0.5}{5.3}{0}{neg}{pos}{\opacity}{0} & \midpointgradientcell{-13.6}{-14.3}{-1.0}{0}{neg}{pos}{\opacity}{0} & \midpointgradientcell{-0.1}{-0.8}{1.7}{0}{neg}{pos}{\opacity}{0} & \midpointgradientcell{-0.2}{-0.7}{0.7}{0}{neg}{pos}{\opacity}{0} & \midpointgradientcell{-4.2}{-7.3}{-0.4}{0}{neg}{pos}{\opacity}{0} \\
 & Q+D LM PRF & \midpointgradientcell{7.3}{1.6}{7.4}{0}{neg}{pos}{\opacity}{1} & \midpointgradientcell{4.6}{-0.4}{4.7}{0}{neg}{pos}{\opacity}{1} & \midpointgradientcell{-8.4}{-15.0}{-3.1}{0}{neg}{pos}{\opacity}{0} & \midpointgradientcell{7.9}{1.9}{11.0}{0}{neg}{pos}{\opacity}{0} & \midpointgradientcell{3.5}{0.5}{5.3}{0}{neg}{pos}{\opacity}{0} & \midpointgradientcell{-6.4}{-14.3}{-1.0}{0}{neg}{pos}{\opacity}{0} & \midpointgradientcell{0.2}{-0.8}{1.7}{0}{neg}{pos}{\opacity}{0} & \midpointgradientcell{0.0}{-0.7}{0.7}{0}{neg}{pos}{\opacity}{0} & \midpointgradientcell{-2.8}{-7.3}{-0.4}{0}{neg}{pos}{\opacity}{0} \\
 \bottomrule
\end{tabular}
}
\caption{How different expansions affect results on datasets that measure \textbf{Domain Shift}. Colors indicate a \hlpos{positive} or \hlneg{negative} delta over \hlbase{scores for no expansion}. Notice that models with higher base scores are generally harmed by expansions while weaker models benefit from them.\vspace{-0.5em}}
\label{tab:domain_shift}
\end{table*}

 \begin{table*}[htb!]
\centering
\small
\begin{tabular}{ll|ccc|ccc}
\toprule
\multicolumn{2}{l}{} & \multicolumn{3}{c|}{\textbf{
Touche-2020}} & \multicolumn{3}{c}{\textbf{Scifact-Refute}}  \\
\scriptsize \textbf{Type} & \scriptsize\textbf{Model} & \scriptsize DPR & \scriptsize Contriever FT & \scriptsize MonoT5-3B  & \scriptsize DPR & \scriptsize Contriever FT & \scriptsize MonoT5-3B \\
\midrule
 & \textit{No Expansion} & \gradientcell{23.0}{22.9}{82.2}{low}{high}{\opacity}{0} & \gradientcell{24.8}{22.9}{82.2}{low}{high}{\opacity}{0} & \gradientcell{32.6}{22.9}{82.2}{low}{high}{\opacity}{0} & \gradientcell{33.9}{22.9}{82.2}{low}{high}{\opacity}{0} & \gradientcell{76.4}{22.9}{82.2}{low}{high}{\opacity}{0} & \gradientcell{82.1}{22.9}{82.2}{low}{high}{\opacity}{0} \\
\midrule
\parbox[t]{2mm}{\multirow{3}{*}{\rotatebox[origin=c]{90}{
 \parbox[c]{1cm}{\centering \scriptsize Query }}}} 
 & HyDE & \midpointgradientcell{-0.3}{-0.4}{0.7}{0}{neg}{pos}{\opacity}{0} & \midpointgradientcell{4.8}{-1.3}{5.2}{0}{neg}{pos}{\opacity}{0} & \midpointgradientcell{-5.9}{-10.2}{-0.8}{0}{neg}{pos}{\opacity}{0} & \midpointgradientcell{-9.1}{-9.2}{9.2}{0}{neg}{pos}{\opacity}{0} & \midpointgradientcell{-0.9}{-4.7}{1.4}{0}{neg}{pos}{\opacity}{0} & \midpointgradientcell{-12.3}{-19.4}{1.0}{0}{neg}{pos}{\opacity}{0} \\
 & CoT & \midpointgradientcell{0.3}{-0.4}{0.7}{0}{neg}{pos}{\opacity}{0} & \midpointgradientcell{5.1}{-1.3}{5.2}{0}{neg}{pos}{\opacity}{1} & \midpointgradientcell{-7.4}{-10.2}{-0.8}{0}{neg}{pos}{\opacity}{0} & \midpointgradientcell{-7.6}{-9.2}{9.2}{0}{neg}{pos}{\opacity}{0} & \midpointgradientcell{0.3}{-4.7}{1.4}{0}{neg}{pos}{\opacity}{0} & \midpointgradientcell{-8.8}{-19.4}{1.0}{0}{neg}{pos}{\opacity}{0} \\
 & Q-LM PRF & \midpointgradientcell{0.6}{-0.4}{0.7}{0}{neg}{pos}{\opacity}{1} & \midpointgradientcell{3.9}{-1.3}{5.2}{0}{neg}{pos}{\opacity}{0} & \midpointgradientcell{-1.3}{-10.2}{-0.8}{0}{neg}{pos}{\opacity}{0} & \midpointgradientcell{6.5}{-9.2}{9.2}{0}{neg}{pos}{\opacity}{0} & \midpointgradientcell{1.1}{-4.7}{1.4}{0}{neg}{pos}{\opacity}{0} & \midpointgradientcell{-1.7}{-19.4}{1.0}{0}{neg}{pos}{\opacity}{0} \\
\midrule
\parbox[t]{2mm}{\multirow{2}{*}{\rotatebox[origin=c]{90}{
 \parbox[c]{0.5cm}{\centering \scriptsize Doc }}}} 
 & D2Q & \midpointgradientcell{-0.2}{-0.4}{0.7}{0}{neg}{pos}{\opacity}{0} & \midpointgradientcell{0.0}{-1.3}{5.2}{0}{neg}{pos}{\opacity}{0} & \midpointgradientcell{-0.9}{-10.2}{-0.8}{0}{neg}{pos}{\opacity}{1} & \midpointgradientcell{2.0}{-9.2}{9.2}{0}{neg}{pos}{\opacity}{0} & \midpointgradientcell{-1.8}{-4.7}{1.4}{0}{neg}{pos}{\opacity}{0} & \midpointgradientcell{0.9}{-19.4}{5.0}{0}{neg}{pos}{\opacity}{1} \\
 & D-LM PRF & \midpointgradientcell{-0.2}{-0.4}{0.7}{0}{neg}{pos}{\opacity}{0} & \midpointgradientcell{-1.2}{-1.3}{5.2}{0}{neg}{pos}{\opacity}{0} & \midpointgradientcell{-8.3}{-10.2}{-0.8}{0}{neg}{pos}{\opacity}{0} & \midpointgradientcell{2.5}{-9.2}{9.2}{0}{neg}{pos}{\opacity}{0} & \midpointgradientcell{-4.6}{-4.7}{1.4}{0}{neg}{pos}{\opacity}{0} & \midpointgradientcell{-16.5}{-19.4}{1.0}{0}{neg}{pos}{\opacity}{0} \\
\midrule
\parbox[t]{2mm}{\multirow{6}{*}{\rotatebox[origin=c]{90}{
 \parbox[c]{1cm}{\centering \scriptsize Both }}}} 
 & HyDE + D2Q & \midpointgradientcell{-0.1}{-0.4}{0.7}{0}{neg}{pos}{\opacity}{0} & \midpointgradientcell{5.0}{-1.3}{5.2}{0}{neg}{pos}{\opacity}{0} & \midpointgradientcell{-3.0}{-10.2}{-0.8}{0}{neg}{pos}{\opacity}{0} & \midpointgradientcell{-6.1}{-9.2}{9.2}{0}{neg}{pos}{\opacity}{0} & \midpointgradientcell{-1.0}{-4.7}{1.4}{0}{neg}{pos}{\opacity}{0} & \midpointgradientcell{-16.6}{-19.4}{1.0}{0}{neg}{pos}{\opacity}{0} \\
 & CoT + D2Q & \midpointgradientcell{0.3}{-0.4}{0.7}{0}{neg}{pos}{\opacity}{0} & \midpointgradientcell{2.6}{-1.3}{5.2}{0}{neg}{pos}{\opacity}{0} & \midpointgradientcell{-5.4}{-10.2}{-0.8}{0}{neg}{pos}{\opacity}{0} & \midpointgradientcell{-6.5}{-9.2}{9.2}{0}{neg}{pos}{\opacity}{0} & \midpointgradientcell{-1.1}{-4.7}{1.4}{0}{neg}{pos}{\opacity}{0} & \midpointgradientcell{-16.9}{-19.4}{1.0}{0}{neg}{pos}{\opacity}{0} \\
 & Q-LM PRF + D2Q & \midpointgradientcell{-0.1}{-0.4}{0.7}{0}{neg}{pos}{\opacity}{0} & \midpointgradientcell{1.0}{-1.3}{5.2}{0}{neg}{pos}{\opacity}{0} & \midpointgradientcell{-2.0}{-10.2}{-0.8}{0}{neg}{pos}{\opacity}{0} & \midpointgradientcell{9.1}{-9.2}{9.2}{0}{neg}{pos}{\opacity}{1} & \midpointgradientcell{1.3}{-4.7}{1.4}{0}{neg}{pos}{\opacity}{1} & \midpointgradientcell{-1.1}{-19.4}{1.0}{0}{neg}{pos}{\opacity}{0} \\
 & HyDE + D-LM PRF & \midpointgradientcell{0.5}{-0.4}{0.7}{0}{neg}{pos}{\opacity}{0} & \midpointgradientcell{1.4}{-1.3}{5.2}{0}{neg}{pos}{\opacity}{0} & \midpointgradientcell{-10.1}{-10.2}{-0.8}{0}{neg}{pos}{\opacity}{0} & \midpointgradientcell{-5.2}{-9.2}{9.2}{0}{neg}{pos}{\opacity}{0} & \midpointgradientcell{-2.9}{-4.7}{1.4}{0}{neg}{pos}{\opacity}{0} & \midpointgradientcell{-17.6}{-19.4}{1.0}{0}{neg}{pos}{\opacity}{0} \\
 & CoT + D-LM PRF & \midpointgradientcell{-0.2}{-0.4}{0.7}{0}{neg}{pos}{\opacity}{0} & \midpointgradientcell{0.8}{-1.3}{5.2}{0}{neg}{pos}{\opacity}{0} & \midpointgradientcell{-8.4}{-10.2}{-0.8}{0}{neg}{pos}{\opacity}{0} & \midpointgradientcell{-7.2}{-9.2}{9.2}{0}{neg}{pos}{\opacity}{0} & \midpointgradientcell{-1.5}{-4.7}{1.4}{0}{neg}{pos}{\opacity}{0} & \midpointgradientcell{-19.3}{-19.4}{1.0}{0}{neg}{pos}{\opacity}{0} \\
 & Q+D LM PRF & \midpointgradientcell{0.3}{-0.4}{0.7}{0}{neg}{pos}{\opacity}{0} & \midpointgradientcell{2.5}{-1.3}{5.2}{0}{neg}{pos}{\opacity}{0} & \midpointgradientcell{-2.7}{-10.2}{-0.8}{0}{neg}{pos}{\opacity}{0} & \midpointgradientcell{7.6}{-9.2}{9.2}{0}{neg}{pos}{\opacity}{0} & \midpointgradientcell{-2.5}{-4.7}{1.4}{0}{neg}{pos}{\opacity}{0} & \midpointgradientcell{-4.0}{-19.4}{1.0}{0}{neg}{pos}{\opacity}{0} \\
\bottomrule
\end{tabular}
\caption{How different expansions affect results on datasets that measure \textbf{Relevance Shift}. 
\vspace{-0.5em}}
\label{tab:relevance_shift}
\end{table*}

Due to the exponential combination of models and datasets, we evaluate all models on three representative datasets in \autoref{tab:models} (we provide a comprehensive description of all datasets in \autoref{sec:rq3});
then, we use five representative models (DPR, Contriever, ColBERTv2, MonoT5-small, and MonoT5-3B) on a larger suite of datasets (see~\autoref{fig:models}).

We present results for expansion technique as absolute increase/decrease in nDCG@10\footnote{Traditional expansion techniques increase  recall of retrieval systems.
However, LM-based expansions have been shown to also improve  precision~\cite{Jagerman2023QueryEB}. 
Thus, we use the official, precision-oriented metric for BEIR, nDCG.} points over a baseline with no expansion, which we highlight in \hlbase{grey} in all tables.
Values above zero (e.g. greater than the base version) are highlighted \hlpos{blue} while values below the base are highlighted \hlneg{red}. 
Color intensity is scaled linearly according to the difference between the base value and the min/max (\textit{i.e.}, more saturation for the highest/lowest values).

We use default hyperparameters for all models, except for the length of the queries, which we set at 512 for BERT-based models and 1024 for T5 and Llama models.

\vspace{-0.2em}
\paragraph{Effect of Different Models}
Our results with all models (\autoref{fig:teaser}) show a consistent pattern: as base performance on a task increases, the gains from expansion decrease. We also see this trend from \autoref{tab:models} (note that ArguAna and FIQA results are sorted by nDCG score on MS MARCO; negative trend is clearly observable in \autoref{fig:teaser}). Interestingly, these results do not depend on the model architecture: this is true for bi-encoders, late-interaction models, neural sparse models, and cross-encoders (of all sizes).

However, do these results hold for other datasets? In \autoref{fig:models}, we show
that this pattern is consistent over a wide range of datasets. Models whose base score is higher (such as MonoT5-3B) are negatively impacted by expansions.

\begin{table*}[t!]
\centering
\small
\resizebox{\textwidth}{!}{\begin{tabular}{ll|ccc|ccc|ccc}
\toprule
\multicolumn{2}{l}{} & \multicolumn{3}{c|}{\textbf{
Tip of My Tongue}} & \multicolumn{3}{c}{\textbf{TREC CT 2021}} & \multicolumn{3}{c}{\textbf{Arguana}}  \\
\scriptsize \textbf{Type} & \scriptsize\textbf{Model} & \scriptsize DPR  & \scriptsize Contriever FT & \scriptsize MonoT5-3B  & \scriptsize DPR  & \scriptsize Contriever FT & \scriptsize MonoT5-3B & \scriptsize DPR  & \scriptsize Contriever FT & \scriptsize MonoT5-3B \\
\midrule
 & \textit{No Expansion} & \gradientcell{13.4}{13.3}{48.9}{low}{high}{\opacity}{0} & \gradientcell{38.3}{13.3}{48.9}{low}{high}{\opacity}{0} & \gradientcell{39.5}{13.3}{48.9}{low}{high}{\opacity}{0} & \gradientcell{16.4}{13.3}{48.9}{low}{high}{\opacity}{0} & \gradientcell{26.7}{13.3}{48.9}{low}{high}{\opacity}{0} & \gradientcell{25.8}{13.3}{48.9}{low}{high}{\opacity}{0} & \gradientcell{34.9}{13.3}{48.9}{low}{high}{\opacity}{0} & \gradientcell{48.8}{13.3}{48.9}{low}{high}{\opacity}{0} & \gradientcell{42.4}{13.3}{48.9}{low}{high}{\opacity}{0} \\
\midrule
\parbox[t]{2mm}{\multirow{3}{*}{\rotatebox[origin=c]{90}{
 \parbox[c]{1cm}{\centering \scriptsize Query }}}} 
 & HyDE & \midpointgradientcell{3.0}{-3.0}{6.1}{0}{neg}{pos}{\opacity}{0} & \midpointgradientcell{-9.4}{-10.8}{3.0}{0}{neg}{pos}{\opacity}{0} & \midpointgradientcell{-26.8}{-32.7}{6.5}{0}{neg}{pos}{\opacity}{0} & \midpointgradientcell{0.3}{-0.8}{2.4}{0}{neg}{pos}{\opacity}{0} & \midpointgradientcell{2.1}{-1.4}{3.1}{0}{neg}{pos}{\opacity}{0} & \midpointgradientcell{4.2}{-1.9}{4.3}{0}{neg}{pos}{\opacity}{1} & \midpointgradientcell{-4.5}{-7.2}{2.4}{0}{neg}{pos}{\opacity}{0} & \midpointgradientcell{-5.4}{-5.5}{3.6}{0}{neg}{pos}{\opacity}{0} & \midpointgradientcell{15.8}{-2.6}{15.9}{0}{neg}{pos}{\opacity}{1} \\
 & CoT & \midpointgradientcell{2.1}{-3.0}{6.1}{0}{neg}{pos}{\opacity}{0} & \midpointgradientcell{-9.5}{-10.8}{3.0}{0}{neg}{pos}{\opacity}{0} & \midpointgradientcell{-23.3}{-32.7}{6.5}{0}{neg}{pos}{\opacity}{0} & \midpointgradientcell{2.3}{-0.8}{2.4}{0}{neg}{pos}{\opacity}{1} & \midpointgradientcell{3.0}{-1.4}{3.1}{0}{neg}{pos}{\opacity}{1} & \midpointgradientcell{3.0}{-1.9}{4.3}{0}{neg}{pos}{\opacity}{0} & \midpointgradientcell{-5.8}{-7.2}{2.4}{0}{neg}{pos}{\opacity}{0} & \midpointgradientcell{-5.3}{-5.5}{3.6}{0}{neg}{pos}{\opacity}{0} & \midpointgradientcell{11.3}{-2.6}{15.9}{0}{neg}{pos}{\opacity}{0} \\
 & Q-LM PRF & \midpointgradientcell{-2.9}{-3.0}{6.1}{0}{neg}{pos}{\opacity}{0} & \midpointgradientcell{-1.9}{-10.8}{3.0}{0}{neg}{pos}{\opacity}{0} & \midpointgradientcell{6.4}{-32.7}{6.5}{0}{neg}{pos}{\opacity}{1} & \midpointgradientcell{2.2}{-0.8}{2.4}{0}{neg}{pos}{\opacity}{0} & \midpointgradientcell{0.6}{-1.4}{3.1}{0}{neg}{pos}{\opacity}{0} & \midpointgradientcell{-0.1}{-1.9}{4.3}{0}{neg}{pos}{\opacity}{0} & \midpointgradientcell{-7.1}{-7.2}{2.4}{0}{neg}{pos}{\opacity}{0} & \midpointgradientcell{-3.6}{-5.5}{3.6}{0}{neg}{pos}{\opacity}{0} & \midpointgradientcell{8.3}{-2.6}{15.9}{0}{neg}{pos}{\opacity}{0} \\
\midrule
\parbox[t]{2mm}{\multirow{2}{*}{\rotatebox[origin=c]{90}{
 \parbox[c]{0.5cm}{\centering \scriptsize Doc }}}} 
 & D2Q & \midpointgradientcell{1.6}{-3.0}{6.1}{0}{neg}{pos}{\opacity}{0} & \midpointgradientcell{-3.2}{-10.8}{3.0}{0}{neg}{pos}{\opacity}{0} & \midpointgradientcell{-8.5}{-32.7}{6.5}{0}{neg}{pos}{\opacity}{0} & \midpointgradientcell{0.3}{-0.8}{2.4}{0}{neg}{pos}{\opacity}{0} & \midpointgradientcell{-1.3}{-1.4}{3.1}{0}{neg}{pos}{\opacity}{0} & \midpointgradientcell{-1.8}{-1.9}{4.3}{0}{neg}{pos}{\opacity}{0} & \midpointgradientcell{1.6}{-7.2}{2.4}{0}{neg}{pos}{\opacity}{0} & \midpointgradientcell{2.0}{-5.5}{3.6}{0}{neg}{pos}{\opacity}{0} & \midpointgradientcell{-2.1}{-2.6}{15.9}{0}{neg}{pos}{\opacity}{0} \\
 & D-LM PRF & \midpointgradientcell{5.5}{-3.0}{6.1}{0}{neg}{pos}{\opacity}{0} & \midpointgradientcell{2.9}{-10.8}{3.0}{0}{neg}{pos}{\opacity}{1} & \midpointgradientcell{0.9}{-32.7}{6.5}{0}{neg}{pos}{\opacity}{0} & \midpointgradientcell{-0.7}{-0.8}{2.4}{0}{neg}{pos}{\opacity}{0} & \midpointgradientcell{-0.9}{-1.4}{3.1}{0}{neg}{pos}{\opacity}{0} & \midpointgradientcell{0.6}{-1.9}{4.3}{0}{neg}{pos}{\opacity}{0} & \midpointgradientcell{2.3}{-7.2}{2.4}{0}{neg}{pos}{\opacity}{1} & \midpointgradientcell{3.5}{-5.5}{3.6}{0}{neg}{pos}{\opacity}{1} & \midpointgradientcell{-2.5}{-2.6}{15.9}{0}{neg}{pos}{\opacity}{0} \\
\midrule
\parbox[t]{2mm}{\multirow{6}{*}{\rotatebox[origin=c]{90}{
 \parbox[c]{1cm}{\centering \scriptsize Both }}}} 
 & HyDE + D2Q & \midpointgradientcell{3.6}{-3.0}{6.1}{0}{neg}{pos}{\opacity}{0} & \midpointgradientcell{-10.7}{-10.8}{3.0}{0}{neg}{pos}{\opacity}{0} & \midpointgradientcell{-29.7}{-32.7}{6.5}{0}{neg}{pos}{\opacity}{0} & \midpointgradientcell{0.4}{-0.8}{2.4}{0}{neg}{pos}{\opacity}{0} & \midpointgradientcell{2.1}{-1.4}{3.1}{0}{neg}{pos}{\opacity}{0} & \midpointgradientcell{2.7}{-1.9}{4.3}{0}{neg}{pos}{\opacity}{0} & \midpointgradientcell{-2.8}{-7.2}{2.4}{0}{neg}{pos}{\opacity}{0} & \midpointgradientcell{-2.5}{-5.5}{3.6}{0}{neg}{pos}{\opacity}{0} & \midpointgradientcell{12.9}{-2.6}{15.9}{0}{neg}{pos}{\opacity}{0} \\
 & CoT + D2Q & \midpointgradientcell{2.2}{-3.0}{6.1}{0}{neg}{pos}{\opacity}{0} & \midpointgradientcell{-10.6}{-10.8}{3.0}{0}{neg}{pos}{\opacity}{0} & \midpointgradientcell{-25.3}{-32.7}{6.5}{0}{neg}{pos}{\opacity}{0} & \midpointgradientcell{2.3}{-0.8}{2.4}{0}{neg}{pos}{\opacity}{1} & \midpointgradientcell{1.5}{-1.4}{3.1}{0}{neg}{pos}{\opacity}{0} & \midpointgradientcell{-0.1}{-1.9}{4.3}{0}{neg}{pos}{\opacity}{0} & \midpointgradientcell{-4.3}{-7.2}{2.4}{0}{neg}{pos}{\opacity}{0} & \midpointgradientcell{-3.0}{-5.5}{3.6}{0}{neg}{pos}{\opacity}{0} & \midpointgradientcell{10.6}{-2.6}{15.9}{0}{neg}{pos}{\opacity}{0} \\
 & Q-LM PRF + D2Q & \midpointgradientcell{-1.8}{-3.0}{6.1}{0}{neg}{pos}{\opacity}{0} & \midpointgradientcell{-4.7}{-10.8}{3.0}{0}{neg}{pos}{\opacity}{0} & \midpointgradientcell{2.1}{-32.7}{6.5}{0}{neg}{pos}{\opacity}{0} & \midpointgradientcell{0.7}{-0.8}{2.4}{0}{neg}{pos}{\opacity}{0} & \midpointgradientcell{-0.9}{-1.4}{3.1}{0}{neg}{pos}{\opacity}{0} & \midpointgradientcell{-0.2}{-1.9}{4.3}{0}{neg}{pos}{\opacity}{0} & \midpointgradientcell{-4.4}{-7.2}{2.4}{0}{neg}{pos}{\opacity}{0} & \midpointgradientcell{-2.5}{-5.5}{3.6}{0}{neg}{pos}{\opacity}{0} & \midpointgradientcell{6.9}{-2.6}{15.9}{0}{neg}{pos}{\opacity}{0} \\
 & HyDE + D-LM PRF & \midpointgradientcell{6.0}{-3.0}{6.1}{0}{neg}{pos}{\opacity}{1} & \midpointgradientcell{-7.2}{-10.8}{3.0}{0}{neg}{pos}{\opacity}{0} & \midpointgradientcell{-32.6}{-32.7}{6.5}{0}{neg}{pos}{\opacity}{0} & \midpointgradientcell{0.0}{-0.8}{2.4}{0}{neg}{pos}{\opacity}{0} & \midpointgradientcell{1.0}{-1.4}{3.1}{0}{neg}{pos}{\opacity}{0} & \midpointgradientcell{3.2}{-1.9}{4.3}{0}{neg}{pos}{\opacity}{0} & \midpointgradientcell{-3.0}{-7.2}{2.4}{0}{neg}{pos}{\opacity}{0} & \midpointgradientcell{1.0}{-5.5}{3.6}{0}{neg}{pos}{\opacity}{0} & \midpointgradientcell{10.3}{-2.6}{15.9}{0}{neg}{pos}{\opacity}{0} \\
 & CoT + D-LM PRF & \midpointgradientcell{5.3}{-3.0}{6.1}{0}{neg}{pos}{\opacity}{0} & \midpointgradientcell{-7.4}{-10.8}{3.0}{0}{neg}{pos}{\opacity}{0} & \midpointgradientcell{-25.8}{-32.7}{6.5}{0}{neg}{pos}{\opacity}{0} & \midpointgradientcell{1.9}{-0.8}{2.4}{0}{neg}{pos}{\opacity}{0} & \midpointgradientcell{2.7}{-1.4}{3.1}{0}{neg}{pos}{\opacity}{0} & \midpointgradientcell{1.0}{-1.9}{4.3}{0}{neg}{pos}{\opacity}{0} & \midpointgradientcell{-4.0}{-7.2}{2.4}{0}{neg}{pos}{\opacity}{0} & \midpointgradientcell{0.9}{-5.5}{3.6}{0}{neg}{pos}{\opacity}{0} & \midpointgradientcell{8.8}{-2.6}{15.9}{0}{neg}{pos}{\opacity}{0} \\
 & Q+D LM PRF & \midpointgradientcell{0.7}{-3.0}{6.1}{0}{neg}{pos}{\opacity}{0} & \midpointgradientcell{1.6}{-10.8}{3.0}{0}{neg}{pos}{\opacity}{0} & \midpointgradientcell{6.4}{-32.7}{6.5}{0}{neg}{pos}{\opacity}{1} & \midpointgradientcell{0.6}{-0.8}{2.4}{0}{neg}{pos}{\opacity}{0} & \midpointgradientcell{-1.0}{-1.4}{3.1}{0}{neg}{pos}{\opacity}{0} & \midpointgradientcell{0.4}{-1.9}{4.3}{0}{neg}{pos}{\opacity}{0} & \midpointgradientcell{-4.0}{-7.2}{2.4}{0}{neg}{pos}{\opacity}{0} & \midpointgradientcell{-0.2}{-5.5}{3.6}{0}{neg}{pos}{\opacity}{0} & \midpointgradientcell{3.3}{-2.6}{15.9}{0}{neg}{pos}{\opacity}{0} \\
\bottomrule
\end{tabular}}
\caption{How different expansions affect results on datasets that measure \textbf{Long Query Format Shift}. Colors indicate a \hlpos{positive} or \hlneg{negative} delta over \hlbase{scores for no expansion}. \textit{Unlike previous results}, all models benefit from expansions on all three datasets. We conclude that, in the case of significant query shift, expansion is useful.\vspace{-0.5em}}
\label{tab:format_shift_query}
\end{table*}

\begin{table*}[htb!]
\centering
\small
\begin{tabular}{ll|ccc|ccc}
\toprule
\multicolumn{2}{l}{} & \multicolumn{3}{c|}{\textbf{
WikiQA}} & \multicolumn{3}{c}{\textbf{Quora}}  \\
\scriptsize \textbf{Type} & \scriptsize\textbf{Model} & \scriptsize DPR & \scriptsize Contriever FT & \scriptsize MonoT5-3B  & \scriptsize DPR & \scriptsize Contriever FT & \scriptsize MonoT5-3B \\
\midrule
 & \textit{No Expansion} & \gradientcell{47.2}{47.1}{86.8}{low}{high}{\opacity}{0} & \gradientcell{68.6}{47.1}{86.8}{low}{high}{\opacity}{0} & \gradientcell{75.9}{47.1}{86.8}{low}{high}{\opacity}{0} & \gradientcell{68.4}{47.1}{86.8}{low}{high}{\opacity}{0} & \gradientcell{86.7}{47.1}{86.8}{low}{high}{\opacity}{0} & \gradientcell{83.9}{47.1}{86.8}{low}{high}{\opacity}{0} \\
\midrule
\parbox[t]{2mm}{\multirow{3}{*}{\rotatebox[origin=c]{90}{
 \parbox[c]{1cm}{\centering \scriptsize Query }}}} 
 & HyDE & \midpointgradientcell{16.4}{-2.9}{17.8}{0}{neg}{pos}{\opacity}{0} & \midpointgradientcell{3.6}{-10.9}{3.7}{0}{neg}{pos}{\opacity}{1} & \midpointgradientcell{-1.6}{-21.5}{-1.5}{0}{neg}{pos}{\opacity}{1} & \midpointgradientcell{-15.4}{-35.7}{-6.1}{0}{neg}{pos}{\opacity}{0} & \midpointgradientcell{-13.8}{-36.9}{-3.6}{0}{neg}{pos}{\opacity}{0} & \midpointgradientcell{-8.2}{-41.5}{0.1}{0}{neg}{pos}{\opacity}{0} \\
 & CoT & \midpointgradientcell{9.8}{-2.9}{17.8}{0}{neg}{pos}{\opacity}{0} & \midpointgradientcell{-0.9}{-10.9}{3.7}{0}{neg}{pos}{\opacity}{0} & \midpointgradientcell{-6.1}{-21.5}{-1.5}{0}{neg}{pos}{\opacity}{0} & \midpointgradientcell{-32.3}{-35.7}{-6.1}{0}{neg}{pos}{\opacity}{0} & \midpointgradientcell{-31.5}{-36.9}{-3.6}{0}{neg}{pos}{\opacity}{0} & \midpointgradientcell{-35.4}{-41.5}{0.1}{0}{neg}{pos}{\opacity}{0} \\
 & Q-LM PRF & \midpointgradientcell{11.9}{-2.9}{17.8}{0}{neg}{pos}{\opacity}{0} & \midpointgradientcell{-2.2}{-10.9}{3.7}{0}{neg}{pos}{\opacity}{0} & \midpointgradientcell{-4.2}{-21.5}{-1.5}{0}{neg}{pos}{\opacity}{0} & \midpointgradientcell{-13.8}{-35.7}{-6.1}{0}{neg}{pos}{\opacity}{0} & \midpointgradientcell{-11.4}{-36.9}{-3.6}{0}{neg}{pos}{\opacity}{0} & \midpointgradientcell{-7.0}{-41.5}{0.1}{0}{neg}{pos}{\opacity}{0} \\
\midrule
\parbox[t]{2mm}{\multirow{2}{*}{\rotatebox[origin=c]{90}{
 \parbox[c]{0.5cm}{\centering \scriptsize Doc }}}} 
 & D2Q & \midpointgradientcell{5.4}{-2.9}{17.8}{0}{neg}{pos}{\opacity}{0} & \midpointgradientcell{-1.8}{-10.9}{3.7}{0}{neg}{pos}{\opacity}{0} & \midpointgradientcell{-1.7}{-21.5}{-1.5}{0}{neg}{pos}{\opacity}{0} & \midpointgradientcell{-6.2}{-35.7}{-6.1}{0}{neg}{pos}{\opacity}{1} & \midpointgradientcell{-3.7}{-36.9}{-3.6}{0}{neg}{pos}{\opacity}{1} & \midpointgradientcell{0.0}{-41.5}{0.1}{0}{neg}{pos}{\opacity}{1} \\
 & D-LM PRF & \midpointgradientcell{-2.8}{-2.9}{17.8}{0}{neg}{pos}{\opacity}{0} & \midpointgradientcell{-10.8}{-10.9}{3.7}{0}{neg}{pos}{\opacity}{0} & \midpointgradientcell{-21.4}{-21.5}{-1.5}{0}{neg}{pos}{\opacity}{0} & \midpointgradientcell{-10.0}{-35.7}{-6.1}{0}{neg}{pos}{\opacity}{0} & \midpointgradientcell{-15.6}{-36.9}{-3.6}{0}{neg}{pos}{\opacity}{0} & \midpointgradientcell{-17.0}{-41.5}{0.1}{0}{neg}{pos}{\opacity}{0} \\
\midrule
\parbox[t]{2mm}{\multirow{6}{*}{\rotatebox[origin=c]{90}{
 \parbox[c]{1cm}{\centering \scriptsize Both }}}} 
 & HyDE + D2Q & \midpointgradientcell{17.7}{-2.9}{17.8}{0}{neg}{pos}{\opacity}{1} & \midpointgradientcell{2.1}{-10.9}{3.7}{0}{neg}{pos}{\opacity}{0} & \midpointgradientcell{-2.7}{-21.5}{-1.5}{0}{neg}{pos}{\opacity}{0} & \midpointgradientcell{-11.4}{-35.7}{-6.1}{0}{neg}{pos}{\opacity}{0} & \midpointgradientcell{-10.1}{-36.9}{-3.6}{0}{neg}{pos}{\opacity}{0} & \midpointgradientcell{-7.1}{-41.5}{0.1}{0}{neg}{pos}{\opacity}{0} \\
 & CoT + D2Q & \midpointgradientcell{11.3}{-2.9}{17.8}{0}{neg}{pos}{\opacity}{0} & \midpointgradientcell{-1.5}{-10.9}{3.7}{0}{neg}{pos}{\opacity}{0} & \midpointgradientcell{-6.9}{-21.5}{-1.5}{0}{neg}{pos}{\opacity}{0} & \midpointgradientcell{-25.7}{-35.7}{-6.1}{0}{neg}{pos}{\opacity}{0} & \midpointgradientcell{-26.3}{-36.9}{-3.6}{0}{neg}{pos}{\opacity}{0} & \midpointgradientcell{-32.5}{-41.5}{0.1}{0}{neg}{pos}{\opacity}{0} \\
 & Q-LM PRF + D2Q & \midpointgradientcell{13.0}{-2.9}{17.8}{0}{neg}{pos}{\opacity}{0} & \midpointgradientcell{-1.1}{-10.9}{3.7}{0}{neg}{pos}{\opacity}{0} & \midpointgradientcell{-6.2}{-21.5}{-1.5}{0}{neg}{pos}{\opacity}{0} & \midpointgradientcell{-9.4}{-35.7}{-6.1}{0}{neg}{pos}{\opacity}{0} & \midpointgradientcell{-8.7}{-36.9}{-3.6}{0}{neg}{pos}{\opacity}{0} & \midpointgradientcell{-6.9}{-41.5}{0.1}{0}{neg}{pos}{\opacity}{0} \\
 & HyDE + D-LM PRF & \midpointgradientcell{12.6}{-2.9}{17.8}{0}{neg}{pos}{\opacity}{0} & \midpointgradientcell{-6.2}{-10.9}{3.7}{0}{neg}{pos}{\opacity}{0} & \midpointgradientcell{-18.0}{-21.5}{-1.5}{0}{neg}{pos}{\opacity}{0} & \midpointgradientcell{-21.1}{-35.7}{-6.1}{0}{neg}{pos}{\opacity}{0} & \midpointgradientcell{-22.1}{-36.9}{-3.6}{0}{neg}{pos}{\opacity}{0} & \midpointgradientcell{-20.2}{-41.5}{0.1}{0}{neg}{pos}{\opacity}{0} \\
 & CoT + D-LM PRF & \midpointgradientcell{7.0}{-2.9}{17.8}{0}{neg}{pos}{\opacity}{0} & \midpointgradientcell{-10.3}{-10.9}{3.7}{0}{neg}{pos}{\opacity}{0} & \midpointgradientcell{-19.0}{-21.5}{-1.5}{0}{neg}{pos}{\opacity}{0} & \midpointgradientcell{-35.6}{-35.7}{-6.1}{0}{neg}{pos}{\opacity}{0} & \midpointgradientcell{-36.8}{-36.9}{-3.6}{0}{neg}{pos}{\opacity}{0} & \midpointgradientcell{-41.4}{-41.5}{0.1}{0}{neg}{pos}{\opacity}{0} \\
 & Q+D LM PRF & \midpointgradientcell{9.5}{-2.9}{17.8}{0}{neg}{pos}{\opacity}{0} & \midpointgradientcell{-6.1}{-10.9}{3.7}{0}{neg}{pos}{\opacity}{0} & \midpointgradientcell{-10.8}{-21.5}{-1.5}{0}{neg}{pos}{\opacity}{0} & \midpointgradientcell{-19.4}{-35.7}{-6.1}{0}{neg}{pos}{\opacity}{0} & \midpointgradientcell{-19.6}{-36.9}{-3.6}{0}{neg}{pos}{\opacity}{0} & \midpointgradientcell{-17.8}{-41.5}{0.1}{0}{neg}{pos}{\opacity}{0} \\
\bottomrule
\end{tabular}
\caption{How different expansions affect results on datasets that measure \textbf{Short Document Format Shift}. Models with higher base scores are generally harmed by expansions while weaker models benefit from them.\vspace{-1.0em}}
\label{tab:format_shift_doc}
\end{table*}

\section{RQ2: How Do Different Distribution Shifts Impact Results?}
\label{sec:rq2}
\paragraph{Experimental Setting}
We evaluate how query and document expansion are impacted by different distribution shifts: in-domain/no shift (MS MARCO), domain shift (e.g. medical, code, legal), relevance shift (finding the opposite or a counterargument), and format shift (extremely long queries or very short documents). Datasets and their descriptive statistics are in~\autoref{tab:dataset_stats}. We use three representative models for these experiments.

\begin{figure*}[t]
    \centering
    \includegraphics[width=0.99\columnwidth,trim=0cm 0cm 0.0cm 0cm]{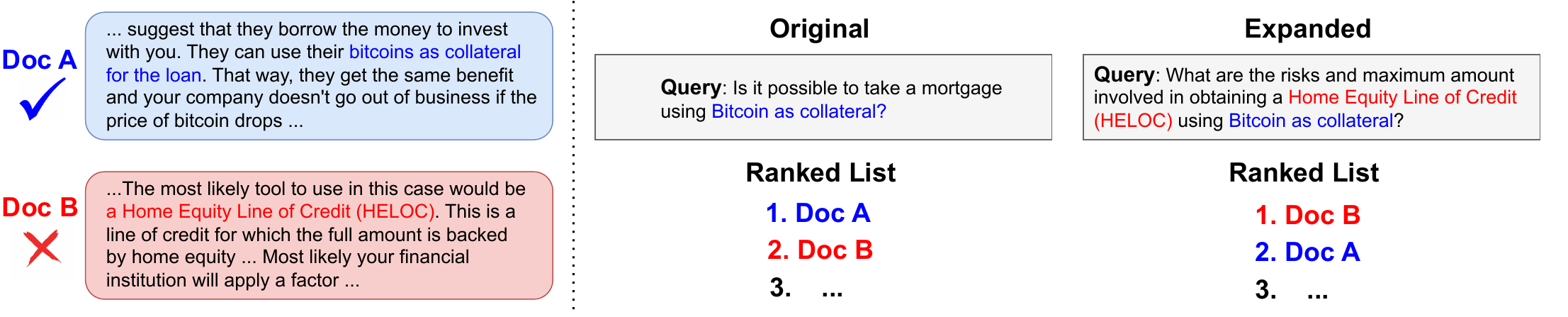}

    \caption{An example of expansions obscuring the relevance signal. The non-relevant document in \textcolor{red}{red (\texttimes )} was ranked higher than the relevant \textcolor{blue}{blue (\checkmark)} document due to the phrase \textcolor{red}{\textit{``Home Equity Line of Credit''}} being added to the query. The left side shows the original query and documents while the right side shows the ranking.\vspace{-1.0em}}
    \label{fig:error_analysis}
\end{figure*}

\paragraph{In-Domain} We use two datasets that test performance on the MS MARCO collection: TREC Deep Learning\footnote{Despite the different names, TREC DL 2019 and 2020 use the same document collection as MS MARCO, albeit with new queries and relevance judgements.} 2019 and 2020 tracks~\cite{craswell2020overview,craswell2021overview}.
All retrieval models considered train on MS MARCO, hence these are \textit{in-domain}.

\paragraph{Domain Shift} In this setting models must generalize from training domain (web documents from MS MARCO) to new domains, such as legal or medical text. This type of shift is made difficult by specialized vocabulary in these domains. We use NFCorpus (medical) \cite{boteva2016}, GooAQ Technical (code) \cite{khashabi2021gooaq}, and FiQA-2018 (finance) \cite{10.1145/3184558.3192301}.

\paragraph{Relevance Shift} This setting is characterized by a difference in how \textit{relevance} is defined. 
Rather than topical relevance over web pages, queries in these datasets ask for counterarguments or documents refuting its claim.
We use two datasets that search for refutations or counterarguments: Touch\'e-2020 \cite{stein:2020v} and a subset of SciFact \cite{wadden-etal-2020-fact} whose gold documents refute the queries claims.

\paragraph{Format Shift} Another type of shift is the length of inputs: generally, queries are short and documents span over one to multiple paragraphs. However, there are situations where queries could be document-sized or the documents could be short. This shift tests whether models can generalize to new length formats. 
We consider two sets of datasets: for \textit{shift to long query} we use the ``Tip of My Tongue'' dataset introduced by ~\citet{lin2023decomposing}, TREC Clinical Trials Track 2021 \cite{roberts2021overview}, and ArguAna \cite{wachsmuth:2018a}. For \textit{shift to short document}, we use Quora \cite{iyer2017quora} and WikiQA \cite{yang2015wikiqa}.

\subsection{Results by Type of Shift}
\autoref{tab:msmarco} shows results for in-domain data on the 2019 and 2020 Deep Learning TREC Tracks. We see that weaker models improve with different expansion types, with DPR improving for almost every expansion and the stronger Contriever showing minor improvements for some combinations. However, when we move to the stronger models (\textit{e.g.}, MonoT5-3B), we find that all of these gains disappear and expansions hurt the model.

We find that this trend holds in most other categories of shift: 
\autoref{tab:domain_shift} for domain shift, \autoref{tab:relevance_shift} for relevance shift, and \autoref{tab:format_shift_doc} for short document shift. Note that \autoref{fig:models} also shows this visually.

The exceptions to this pattern occur in format shift: on Quora (\autoref{tab:relevance_shift}), all models are harmed by expansion;
for long query shift (\autoref{tab:format_shift_query}), expansions generally help most models. 
When we examine why expansions help for the latter, we find that the transformations typically shorten queries to more closely resemble models' training data (\textit{e.g.}, for ArguAna the query changes from a long document to a shorter sentence that summarizes it).

As IR models are not typically trained on long queries, it is an open-question of whether additional training would make this category of shift easier for models and thus make expansions less helpful.

\section{RQ3: Why Do Expansions Hurt?}
\label{sec:rq3}
Sections \ref{sec:rq1} and \ref{sec:rq2} show that strong IR models do not benefit from expansions. But what causes this effect? 
Here, we explore whether model size (\S\ref{sub:model-size}) is linked to our findings, and perform a qualitative error analysis (\S\ref{sec:error}).

\subsection{
Drop in Performance Independent of Size}
\label{sub:model-size}
One possible argument is that larger models are able to estimate relevance better when using unaltered queries and documents, as they have learned a more refined relevance model during their training.
To verify this hypothesis, we test two different families of models: MonoT5 and E5. 
If model size is the cause, we would expect to see larger models gain less from expansions for both families. 

\begin{figure}[htb]
    \centering
    \includegraphics[width=0.999\columnwidth,trim=0.5cm 2cm 0.0cm -0.25cm]{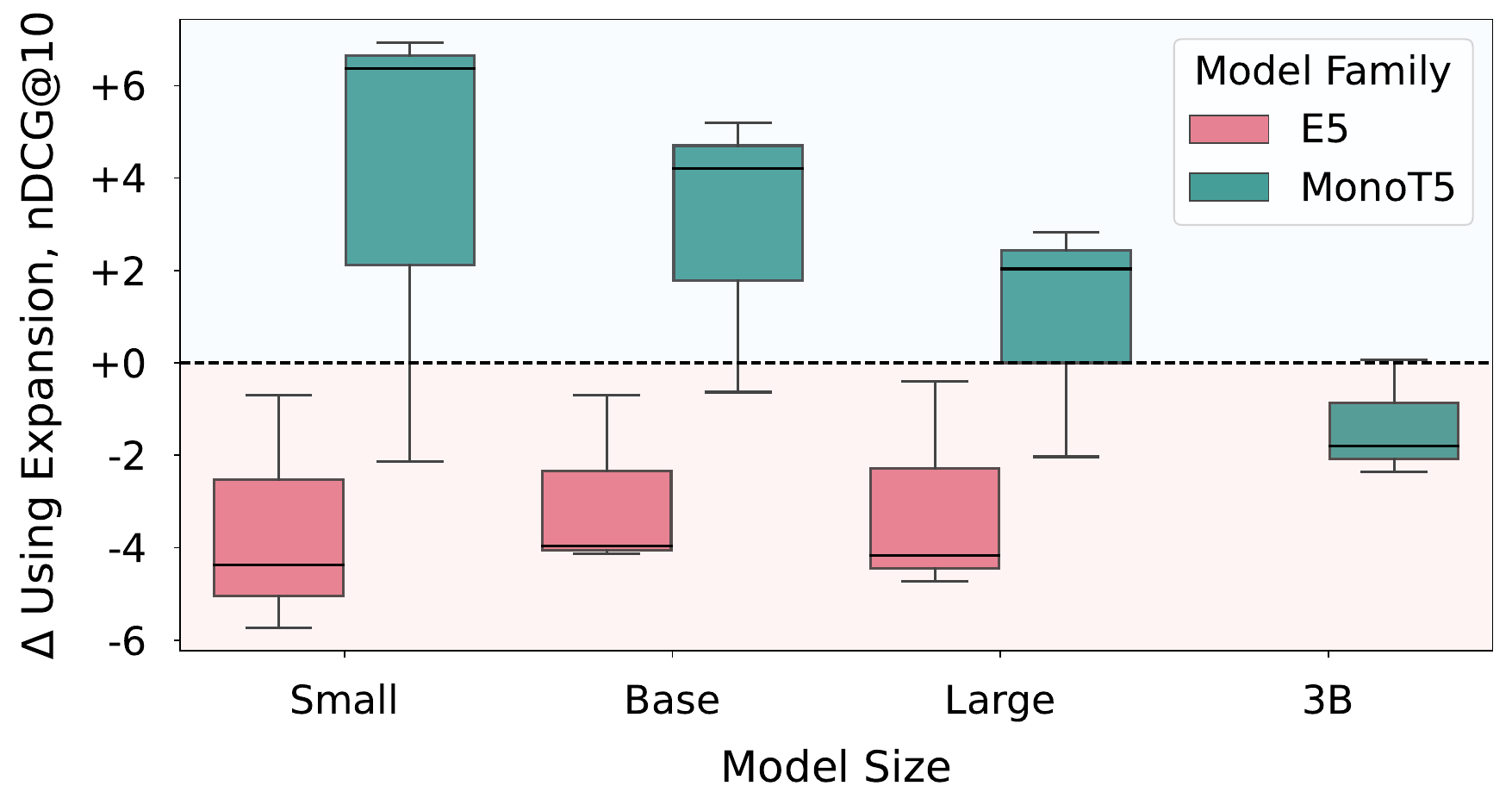}

    \caption{Model scale does \textbf{not} explain negative effect of LM-based expansions. While larger MonoT5 models perform worse, all E5 model sizes are equally impacted}.\vspace{-0.5em}
    \label{fig:scale}
\end{figure}

However, \autoref{fig:scale} shows that model scale is inversely correlated with gains from expansion for the MonoT5-family, but not the E5-family. The crucial difference between them\footnote{Another obvious difference is that E5 is a bi-encoder while MonoT5 is not. However, previous work \cite{muennighoff2022sgpt} has shown that bi-encoders also improve with scale.} can be attributed to the E5 models having similar performance scores across sizes whereas T5 has a much wider range: T5 differs by 21 nDCG@10 points on ArguAna from 3B to small while E5 differs by only 3 points from large to small. Thus, we see that model size impacts gains from expansions only in tandem with the correlation between model size and base score.

\subsection{Error Analysis}
\label{sec:error}
If model size is not the reason for our finding, what could be causing it? 
To gain an intuition on the failures of LM expansion, we annotate 30 examples from three datasets where performance declines when expanding queries and documents.

\begin{figure}[t!]
    \centering
    \includegraphics[width=0.999\columnwidth,trim=0.5cm 0.5cm 0.0cm 0.25cm]{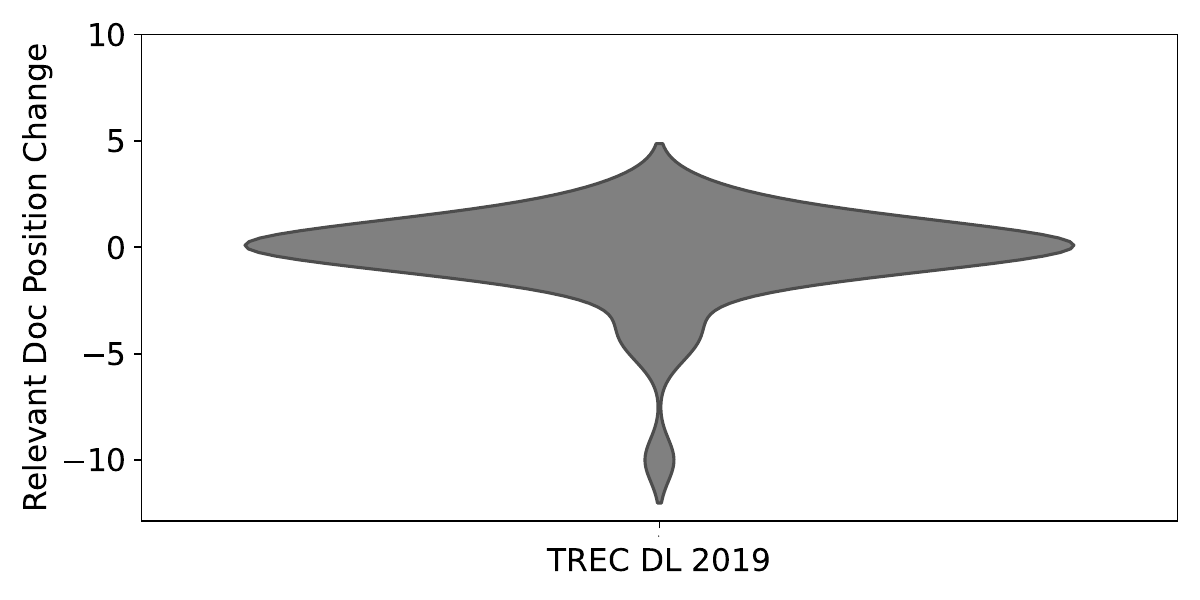}

    \caption{The change in rank for relevant documents in the top 10 when using expansions. Negative values indicate lower ranks (e.g. -5 indicates that the rank of the relevant document went down 5 when using expansions). We see that expansions cause relevant documents to be ranked lower.  \autoref{fig:position_change_app} in the Appendix shows other datasets with similar results.\vspace{-0.5em}}
    \label{fig:position_change}
\end{figure}

We find that out of the 30 examples, two are false negatives, \textit{i.e.}, relevant documents that are unjudged and not labeled as relevant (both from FiQA).
Of the remaining 28, all errors are due to the expansions adding irrelevant terms that dilute relevance signal, or including erroneous keywords that make irrelevant documents appear relevant.
\autoref{fig:error_analysis} shows an example of how query expansion added the term \textit{``Home Equity Line of Credit''} and distracted from the main focus of the question (using bitcoins as collateral).
Thus, it is likely that, without the noise LM-based expansions introduce, well tuned rankers can accurately estimate relevance of subtly different documents. 
We can visualize this in \autoref{fig:position_change}, where we note a general downward shift of the rankings of relevant documents in the top-10 positions for TREC DL 2019. We find that most expansions shifts the ranking by a few positions, while some expansions shift the relevant document ranks to be out of the top 10 (i.e. the cluster at -10 in \autoref{fig:position_change}).

\section{Discussion}
Our results indicate three phenomena regarding expansion using LMs: (\textit{i}) expansion generally benefits weaker models, such as DPR, while better performing rankers, such as T5, are penalized; 
(\textit{ii}) exceptions are observed in case of severe distribution shift, \textit{e.g.}~very long queries; 
(\textit{iii}) when model scores decrease, the cause is generally expansion weakening the original relevance signal.

This implies that despite their broad capabilities, LMs should not be used to augment strong performing IR models without careful testing. 
The strong performance of rerankers for generalization confirms previous work by~\citet{rosa2022defense}.
Further, Table~\ref{tab:msmarco} indicates this characterization of LM expansion also holds on in-domain data (no shift). 

Interestingly, our experiments find that the only distribution shift that consistently needs expansion is long query format shift;
we found no equivalent result for domain, document, or relevance shift. 
Future work may examine whether improved training techniques on longer queries can overcome this or whether longer queries are innately more difficult.

\section{Related Work}

\paragraph{Large Scale Analyses in Neural IR}
Comprehensive analyses in retrieval have provided great insight into practical uses of retrieval. These include many aspects of information retrieval, including interpretability~\cite{macavaney2022abnirml}, domain changes~\cite{simon2023MsShift}, syntax phenomena~\cite{Chari2023OnTE,Weller2023NevIRNI}, and relationship between neural and classical IR approaches~\cite{Formal2021MatchYW,Chen2021SalientPA}.

\paragraph{Generalization in Neural IR} 
As retrieval models have become more effective, attention has turned to improving and evaluating the way that IR models generalize to out-of-distribution datasets (e.g. not MS MARCO-like corpora). One prominent example of this is the BEIR dataset suite \cite{thakur2021beir}, which is commonly used for retrieval evaluation. Much other work has proposed new datasets for types of shift (e.g. MTEB  \cite{muennighoff2022mteb} among others \cite{Han2023RobustQABT,Ravfogel2023RetrievingTB,Weller2023NevIRNI}), as well as many new modeling strategies for better zero-shot retrieval \cite{Dai2022PromptagatorFD,Wang2021GPLGP}. We follow these works by showing different types of shift and whether these types of shift change the results for LM-based expansion techniques.

\paragraph{Effect of Scale on Neural IR Models}
IR models typically improve with scale \cite{nogueira2020document} but are also heavily constrained, due to the requirement of processing  documents for live search. Thus, most first-stage IR models typically use a BERT backbone \cite{santhanam2021colbertv2,izacard2021unsupervised} while reranker models have scaled to billions of parameters \cite{nogueira2020document}. However, work on scaling bi-encoder architectures has also shown performance gains from scale \cite{muennighoff2022sgpt}.  Due to the effectiveness of larger models, recent work has shown that a better first-stage model does not lead to improvements over a BM25 + reranker pipeline \cite{rosa2022defense}. Thus, for our experiments we use BM25 as first stage retrieval and show results reranking those.

\paragraph{Query and Document Expansion in IR}
Query and document expansion have a long history in IR, with early techniques such as expanding query terms using dictionaries or other hand-built knowledge sources \cite{smeaton1995trec,Liu2004AnEA} as well as techniques that use corpus-specific information such as pseudo-relevance feedback \cite{rocchio1971relevance}. These expansions are limited as they are either hand-crafted (and thus limited in scope) or involved automatic techniques that may introduce spurious connections between words. LM-based query and document expansions on the other hand can rely on their extensive linguistic knowledge which goes well beyond hand-crafted rules. Despite this however, they still suffer from spurious and superfluous additions, as shown in \autoref{fig:error_analysis}. However, LM-based expansions have been shown to be successful in a variety of applications \cite{zheng2020bert,weller2022defending,wang2023query2doc,Jagerman2023QueryEB}, which provided inspiration for this work. 

\section{Conclusion}
We conduct the first large scale analysis on large language model (LM) based query and document expansion, studying how model performance, architecture, and size affects these results.
We find that these expansions improve weaker IR models while generally harming performance for the strongest models (including large rerankers and heavily optimized first-stage models).
We further show that this negative correlation between model performance and gains from expansion are true for a wide variety of out of distribution datasets, except for long query shift, where this correlation is weaker.
Overall, our results indicate that LM expansion should not be used for stronger IR models and should instead be confined to weaker retrieval models.

\section*{Limitations}

\paragraph{We evaluate rankers in a zero-shot setup.} This work does not train rankers to deal with augmentations.
While additional training might help mitigate negative effect of document and query expansion, it would significantly increase computational requirements. 
In fact, as our analysis reveals that no single expansion technique is superior in all settings, users would need to train rankers for multiple expansion techniques, further increasing the cost of this fine-tuning step. 
Finally, some tasks might require fine-tuning on supervised data, which might not be available or easily obtainable.

\paragraph{Our protocol for choosing whether a ranker need expansion requires labeled test data in the target domains.}
While our work requires no labeled data to train models, we note that deciding whether to use augmentation requires having access to evaluation data for the target domain: in some cases, such data might not be available.
While recently proposed LM-aided IR evaluation techniques~\cite{Faggioli2023Perspectives,MacAvaney2023OneShot,Thomas2023LargeLM} might ameliorate the need of supervised data, we do not explore such approaches in this work.

\paragraph{While open LMs were evaluated, majority of experiments rely on commercial LM APIs.} 
The majority of experiments in this work were carried out with commercial language models available via paid APIs. 
While we experimented with a variety of other paid API and open LMs (\gptfour, Claude V2, Llama2 70b Chat), we found that they all generally show similar trends, with commercial APIs currently outperforming open models (see Appendix~\ref{app:LMs} and Table~\ref{tab:llms} for more details). 
As our work is mainly focused on studying the effect of expansion different rankers, we feel picking one representative model is justified. 
Nevertheless, use of commercial APIs limits reproducibility and presents a significant barrier to those who cannot get access to the model. 
To minimize this, we will release all LM generations gathered from commercial APIs and from open-source models.

\paragraph{Compute requirements to fully replicate this work.} A replication of this work would require access to significant computational resources, including GPUs. 
A rough estimate shows that generating results for this paper required north of 10,000 NVIDIA A6000 GPU hours, with a further 5,000 hours to develop a stable experimental platform. 

\paragraph{Only English information retrieval models are evaluated.} This work only studies datasets in English. While LM augmentations could play an important role in improving non-English, cross-lingual, and multilingual information retrieval, they require careful analysis (c.f. \citet{mayfield2023synthetic} as one example).

\section*{Ethical Considerations}

\paragraph{LMs may generate factually incorrect text, which could affect ranking.}
This work shows that LM augmentations make mistakes; while our experimental setup is such that LM-generated content never replaces actual documents, inaccuracies might result in non-relevant documents being presented to users.

\section*{Acknowledgements}
OW is supported by the National Science Foundation Graduate Research Fellowship Program. We thanks Sean MacAvaney, Akari Asai, and the Semantic Scholar team at AI2 for their feedback and comments that helped improve this work, as well as those of the reviewers.

\bibliography{anthology,custom}

\appendix

\section{Different LMs for Expansion}
\label{app:LMs}
Here we show results for GPT-4 expansions instead of ChatGPT in \autoref{tab:llms}. We can see that although absolute numbers differ slightly, there is no change to the trends discussed in the main paper: i.e. that stronger models are harmed by expansions while weaker models benefit. We swap FiQA for NFCorpus due to the larger collection size and increased costs of annotating with GPT-4.

\begin{table*}[t!]
\centering
\small
\resizebox{\textwidth}{!}{\begin{tabular}{ll|ccc|ccc|ccc}
\toprule
\multicolumn{2}{l}{} & \multicolumn{3}{c|}{\textbf{
TREC DL 2019}} & \multicolumn{3}{c}{\textbf{NFCorpus}} & \multicolumn{3}{c}{\textbf{Arguana}} \\
\scriptsize \textbf{Type} & \scriptsize\textbf{Model} & \scriptsize DPR  & \scriptsize Contriever FT & \scriptsize MonoT5-3B  & \scriptsize DPR  & \scriptsize Contriever FT & \scriptsize MonoT5-3B & \scriptsize DPR  & \scriptsize Contriever FT & \scriptsize MonoT5-3B \\
\midrule
 & \textit{No Expansion} & \gradientcell{38.4}{24.0}{71.8}{low}{high}{\opacity}{0} & \gradientcell{62.3}{24.0}{71.8}{low}{high}{\opacity}{0} & \gradientcell{71.7}{24.0}{71.8}{low}{high}{\opacity}{0} & \gradientcell{24.1}{24.0}{71.8}{low}{high}{\opacity}{0} & \gradientcell{34.6}{24.0}{71.8}{low}{high}{\opacity}{0} & \gradientcell{39.2}{24.0}{71.8}{low}{high}{\opacity}{0} & \gradientcell{34.9}{24.0}{71.8}{low}{high}{\opacity}{0} & \gradientcell{48.8}{24.0}{71.8}{low}{high}{\opacity}{0} & \gradientcell{42.4}{24.0}{71.8}{low}{high}{\opacity}{0} \\
 
\midrule

\parbox[t]{2mm}{\multirow{3}{*}{\rotatebox[origin=c]{90}{
 \parbox[c]{1cm}{\centering \scriptsize ChatGPT }}}} 
 & Q-LM PRF & \midpointgradientcell{6.6}{3.0}{10.9}{0}{neg}{pos}{\opacity}{0} & \midpointgradientcell{1.6}{-0.3}{1.7}{0}{neg}{pos}{\opacity}{1} & \midpointgradientcell{-2.8}{-5.1}{-1.9}{0}{neg}{pos}{\opacity}{0} & \midpointgradientcell{0.2}{0.1}{1.5}{0}{neg}{pos}{\opacity}{0} & \midpointgradientcell{-0.4}{-0.5}{0.7}{0}{neg}{pos}{\opacity}{0} & \midpointgradientcell{-2.8}{-3.1}{-0.4}{0}{neg}{pos}{\opacity}{0} & \midpointgradientcell{-7.1}{-7.2}{1.7}{0}{neg}{pos}{\opacity}{0} & \midpointgradientcell{-3.6}{-3.7}{2.1}{0}{neg}{pos}{\opacity}{0} & \midpointgradientcell{6.8}{-2.0}{6.9}{0}{neg}{pos}{\opacity}{1} \\
 & D2Q & \midpointgradientcell{3.1}{3.0}{10.9}{0}{neg}{pos}{\opacity}{0} & \midpointgradientcell{-0.2}{-0.3}{1.7}{0}{neg}{pos}{\opacity}{0} & \midpointgradientcell{-2.0}{-5.1}{-1.9}{0}{neg}{pos}{\opacity}{1} & \midpointgradientcell{1.3}{0.1}{1.5}{0}{neg}{pos}{\opacity}{0} & \midpointgradientcell{0.6}{-0.5}{0.7}{0}{neg}{pos}{\opacity}{1} & \midpointgradientcell{-0.5}{-3.1}{-0.4}{0}{neg}{pos}{\opacity}{1} & \midpointgradientcell{1.6}{-7.2}{1.7}{0}{neg}{pos}{\opacity}{1} & \midpointgradientcell{2.0}{-3.7}{2.1}{0}{neg}{pos}{\opacity}{1} & \midpointgradientcell{-1.9}{-2.0}{6.9}{0}{neg}{pos}{\opacity}{0} \\
 & Q-LM PRF + D2Q & \midpointgradientcell{10.8}{3.0}{10.9}{0}{neg}{pos}{\opacity}{1} & \midpointgradientcell{0.6}{-0.3}{1.7}{0}{neg}{pos}{\opacity}{0} & \midpointgradientcell{-5.0}{-5.1}{-1.9}{0}{neg}{pos}{\opacity}{0} & \midpointgradientcell{1.4}{0.1}{1.5}{0}{neg}{pos}{\opacity}{1} & \midpointgradientcell{-0.1}{-0.5}{0.7}{0}{neg}{pos}{\opacity}{0} & \midpointgradientcell{-3.0}{-3.1}{-0.4}{0}{neg}{pos}{\opacity}{0} & \midpointgradientcell{-4.4}{-7.2}{1.7}{0}{neg}{pos}{\opacity}{0} & \midpointgradientcell{-2.5}{-3.7}{2.1}{0}{neg}{pos}{\opacity}{0} & \midpointgradientcell{5.2}{-2.0}{6.9}{0}{neg}{pos}{\opacity}{0} \\
 
 \midrule
 
\parbox[t]{2mm}{\multirow{3}{*}{\rotatebox[origin=c]{90}{
 \parbox[c]{1cm}{\centering \scriptsize GPT-4 }}}} 
 & Q-LM PRF & \midpointgradientcell{13.3}{-4.4}{13.4}{0}{neg}{pos}{\opacity}{1} & \midpointgradientcell{5.2}{-14.1}{5.3}{0}{neg}{pos}{\opacity}{1} & \midpointgradientcell{-0.6}{-3.3}{-0.5}{0}{neg}{pos}{\opacity}{1} & \midpointgradientcell{-7.8}{-7.9}{1.3}{0}{neg}{pos}{\opacity}{0} & \midpointgradientcell{-17.5}{-17.9}{1.1}{0}{neg}{pos}{\opacity}{0} & \midpointgradientcell{-22.6}{-23.4}{0.0}{0}{neg}{pos}{\opacity}{0} & \midpointgradientcell{-6.2}{-6.3}{1.0}{0}{neg}{pos}{\opacity}{0} & \midpointgradientcell{-4.5}{-4.6}{1.3}{0}{neg}{pos}{\opacity}{0} & \midpointgradientcell{4.5}{0.1}{5.3}{0}{neg}{pos}{\opacity}{0} \\
 & D2Q & \midpointgradientcell{-4.3}{-4.4}{13.4}{0}{neg}{pos}{\opacity}{0} & \midpointgradientcell{-14.0}{-14.1}{5.3}{0}{neg}{pos}{\opacity}{0} & \midpointgradientcell{-2.3}{-3.3}{-0.5}{0}{neg}{pos}{\opacity}{0} & \midpointgradientcell{1.2}{-7.9}{1.3}{0}{neg}{pos}{\opacity}{1} & \midpointgradientcell{1.0}{-17.9}{1.1}{0}{neg}{pos}{\opacity}{1} & \midpointgradientcell{-0.1}{-23.4}{0.0}{0}{neg}{pos}{\opacity}{1} & \midpointgradientcell{0.9}{-6.3}{1.0}{0}{neg}{pos}{\opacity}{1} & \midpointgradientcell{1.2}{-4.6}{1.3}{0}{neg}{pos}{\opacity}{1} & \midpointgradientcell{0.2}{0.1}{5.3}{0}{neg}{pos}{\opacity}{0} \\
 & Q-LM PRF + D2Q & \midpointgradientcell{8.0}{-4.4}{13.4}{0}{neg}{pos}{\opacity}{0} & \midpointgradientcell{-8.6}{-14.1}{5.3}{0}{neg}{pos}{\opacity}{0} & \midpointgradientcell{-3.2}{-3.3}{-0.5}{0}{neg}{pos}{\opacity}{0} & \midpointgradientcell{-7.6}{-7.9}{1.3}{0}{neg}{pos}{\opacity}{0} & \midpointgradientcell{-17.8}{-17.9}{1.1}{0}{neg}{pos}{\opacity}{0} & \midpointgradientcell{-23.3}{-23.4}{0.0}{0}{neg}{pos}{\opacity}{0} & \midpointgradientcell{-4.8}{-6.3}{1.0}{0}{neg}{pos}{\opacity}{0} & \midpointgradientcell{-2.9}{-4.6}{1.3}{0}{neg}{pos}{\opacity}{0} & \midpointgradientcell{5.2}{0.1}{5.3}{0}{neg}{pos}{\opacity}{1} \\
\midrule
\parbox[t]{2mm}{\multirow{3}{*}{\rotatebox[origin=c]{90}{
 \parbox[c]{1cm}{\centering \scriptsize Claude v2 }}}} 
 & PRF & \midpointgradientcell{14.0}{4.1}{15.4}{0}{neg}{pos}{\opacity}{0} & \midpointgradientcell{4.8}{-1.8}{4.9}{0}{neg}{pos}{\opacity}{1} & \midpointgradientcell{-3.7}{-4.5}{-2.3}{0}{neg}{pos}{\opacity}{0} & \midpointgradientcell{0.3}{0.2}{1.7}{0}{neg}{pos}{\opacity}{0} & \midpointgradientcell{1.1}{0.4}{1.7}{0}{neg}{pos}{\opacity}{0} & \midpointgradientcell{-1.5}{-1.7}{-0.1}{0}{neg}{pos}{\opacity}{0} & \midpointgradientcell{-6.0}{-6.1}{3.5}{0}{neg}{pos}{\opacity}{0} & \midpointgradientcell{-5.7}{-5.8}{3.4}{0}{neg}{pos}{\opacity}{0} & \midpointgradientcell{4.0}{-1.1}{4.1}{0}{neg}{pos}{\opacity}{1} \\
 & D2Q & \midpointgradientcell{4.2}{4.1}{15.4}{0}{neg}{pos}{\opacity}{0} & \midpointgradientcell{-1.7}{-1.8}{4.9}{0}{neg}{pos}{\opacity}{0} & \midpointgradientcell{-2.4}{-4.5}{-2.3}{0}{neg}{pos}{\opacity}{1} & \midpointgradientcell{1.6}{0.2}{1.7}{0}{neg}{pos}{\opacity}{1} & \midpointgradientcell{0.5}{0.4}{1.7}{0}{neg}{pos}{\opacity}{0} & \midpointgradientcell{-0.2}{-1.7}{-0.1}{0}{neg}{pos}{\opacity}{1} & \midpointgradientcell{3.4}{-6.1}{3.5}{0}{neg}{pos}{\opacity}{1} & \midpointgradientcell{3.3}{-5.8}{3.4}{0}{neg}{pos}{\opacity}{1} & \midpointgradientcell{-1.0}{-1.1}{4.1}{0}{neg}{pos}{\opacity}{0} \\
 & PRF + D2Q & \midpointgradientcell{15.3}{4.1}{15.4}{0}{neg}{pos}{\opacity}{1} & \midpointgradientcell{2.6}{-1.8}{4.9}{0}{neg}{pos}{\opacity}{0} & \midpointgradientcell{-4.4}{-4.5}{-2.3}{0}{neg}{pos}{\opacity}{0} & \midpointgradientcell{1.5}{0.2}{1.7}{0}{neg}{pos}{\opacity}{0} & \midpointgradientcell{1.6}{0.4}{1.7}{0}{neg}{pos}{\opacity}{1} & \midpointgradientcell{-1.6}{-1.7}{-0.1}{0}{neg}{pos}{\opacity}{0} & \midpointgradientcell{-3.1}{-6.1}{3.5}{0}{neg}{pos}{\opacity}{0} & \midpointgradientcell{-2.1}{-5.8}{3.4}{0}{neg}{pos}{\opacity}{0} & \midpointgradientcell{3.7}{-1.1}{4.1}{0}{neg}{pos}{\opacity}{0} \\
\midrule
\parbox[t]{2mm}{\multirow{3}{*}{\rotatebox[origin=c]{90}{
 \parbox[c]{1cm}{\centering \scriptsize Llama v2 70B Chat }}}} 
 & PRF & \midpointgradientcell{0.9}{0.8}{4.8}{0}{neg}{pos}{\opacity}{0} & \midpointgradientcell{-8.3}{-8.4}{-1.0}{0}{neg}{pos}{\opacity}{0} & \midpointgradientcell{-14.5}{-15.9}{-2.4}{0}{neg}{pos}{\opacity}{0} & \midpointgradientcell{-1.5}{-1.6}{1.1}{0}{neg}{pos}{\opacity}{0} & \midpointgradientcell{-1.7}{-1.8}{0.3}{0}{neg}{pos}{\opacity}{0} & \midpointgradientcell{-3.9}{-4.3}{-0.1}{0}{neg}{pos}{\opacity}{0} & \midpointgradientcell{-4.8}{-4.9}{0.0}{0}{neg}{pos}{\opacity}{0} & \midpointgradientcell{-4.5}{-4.6}{1.0}{0}{neg}{pos}{\opacity}{0} & \midpointgradientcell{-2.6}{-4.1}{-2.4}{0}{neg}{pos}{\opacity}{0} \\
 & D2Q & \midpointgradientcell{4.7}{0.8}{4.8}{0}{neg}{pos}{\opacity}{1} & \midpointgradientcell{-1.1}{-8.4}{-1.0}{0}{neg}{pos}{\opacity}{1} & \midpointgradientcell{-2.5}{-15.9}{-2.4}{0}{neg}{pos}{\opacity}{1} & \midpointgradientcell{1.0}{-1.6}{1.1}{0}{neg}{pos}{\opacity}{1} & \midpointgradientcell{0.2}{-1.8}{0.3}{0}{neg}{pos}{\opacity}{1} & \midpointgradientcell{-0.2}{-4.3}{-0.1}{0}{neg}{pos}{\opacity}{1} & \midpointgradientcell{-0.1}{-4.9}{0.0}{0}{neg}{pos}{\opacity}{1} & \midpointgradientcell{0.9}{-4.6}{1.0}{0}{neg}{pos}{\opacity}{1} & \midpointgradientcell{-2.5}{-4.1}{-2.4}{0}{neg}{pos}{\opacity}{1} \\
 & PRF + D2Q & \midpointgradientcell{3.6}{0.8}{4.8}{0}{neg}{pos}{\opacity}{0} & \midpointgradientcell{-7.8}{-8.4}{-1.0}{0}{neg}{pos}{\opacity}{0} & \midpointgradientcell{-15.8}{-15.9}{-2.4}{0}{neg}{pos}{\opacity}{0} & \midpointgradientcell{-0.7}{-1.6}{1.1}{0}{neg}{pos}{\opacity}{0} & \midpointgradientcell{-1.7}{-1.8}{0.3}{0}{neg}{pos}{\opacity}{0} & \midpointgradientcell{-4.2}{-4.3}{-0.1}{0}{neg}{pos}{\opacity}{0} & \midpointgradientcell{-4.3}{-4.9}{0.0}{0}{neg}{pos}{\opacity}{0} & \midpointgradientcell{-3.4}{-4.6}{1.0}{0}{neg}{pos}{\opacity}{0} & \midpointgradientcell{-4.0}{-4.1}{-2.4}{0}{neg}{pos}{\opacity}{0} \\
 
\bottomrule
\end{tabular}}
\caption{How different LLMs used as the generator affect results. Colors indicate a \hlpos{positive} or \hlneg{negative} delta over \hlbase{scores for no expansion}. Although there are small differences \textbf{the overall trends are the same}.\vspace{-0.5em}}
\label{tab:llms}
\end{table*}

\section{Placement of Expansions}
\label{app:placement}
In \autoref{tab:placement} we show different placements of expansions (i.e. do we prepend/append/replace the original query when doing query expansion?). We find that the placement does not make a significant difference to our overall results, as the core conclusion of the paper remains the same.

\begin{table*}[htb!]
\centering
\small
\resizebox{\textwidth}{!}{\begin{tabular}{ll|ccc|ccc|ccc}
\toprule
\multicolumn{2}{l}{} & \multicolumn{3}{c|}{\textbf{
MSMarco 2019}} & \multicolumn{3}{c}{\textbf{FiQA}} & \multicolumn{3}{c}{\textbf{Arguana}} \\
\scriptsize \textbf{Type} & \scriptsize\textbf{Model} & \scriptsize Contriever & \scriptsize MonoT5-small & \scriptsize MonoT5-3B & \scriptsize Contriever & \scriptsize MonoT5-small & \scriptsize MonoT5-3B & \scriptsize Contriever & \scriptsize MonoT5-small & \scriptsize MonoT5-3B\\
\midrule

    & \textit{No Expansion} & \gradientcell{14.4}{14.3}{80.3}{low}{high}{\opacity}{0} & \gradientcell{29.6}{14.3}{80.3}{low}{high}{\opacity}{0} & \gradientcell{45.9}{14.3}{80.3}{low}{high}{\opacity}{0} & \gradientcell{42.5}{14.3}{80.3}{low}{high}{\opacity}{0} & \gradientcell{71.0}{14.3}{80.3}{low}{high}{\opacity}{0} & \gradientcell{80.2}{14.3}{80.3}{low}{high}{\opacity}{0} & \gradientcell{24.1}{14.3}{80.3}{low}{high}{\opacity}{0} & \gradientcell{34.6}{14.3}{80.3}{low}{high}{\opacity}{0} & \gradientcell{39.2}{14.3}{80.3}{low}{high}{\opacity}{0} \\
\midrule
\parbox[t]{2mm}{\multirow{3}{*}{\rotatebox[origin=c]{90}{
 \parbox[c]{1cm}{\centering \scriptsize Query }}}} 
 & Prepend & \midpointgradientcell{8.1}{2.6}{11.1}{0}{neg}{pos}{\opacity}{0} & \midpointgradientcell{-2.8}{-11.9}{0.3}{0}{neg}{pos}{\opacity}{0} & \midpointgradientcell{-4.2}{-8.8}{-0.6}{0}{neg}{pos}{\opacity}{0} & \midpointgradientcell{5.1}{0.8}{8.4}{0}{neg}{pos}{\opacity}{0} & \midpointgradientcell{-0.3}{-8.6}{0.9}{0}{neg}{pos}{\opacity}{0} & \midpointgradientcell{-5.6}{-10.5}{-1.1}{0}{neg}{pos}{\opacity}{0} & \midpointgradientcell{-3.2}{-16.0}{3.0}{0}{neg}{pos}{\opacity}{0} & \midpointgradientcell{22.2}{-11.1}{22.8}{0}{neg}{pos}{\opacity}{0} & \midpointgradientcell{6.9}{-26.9}{8.5}{0}{neg}{pos}{\opacity}{0} \\
 & Append & \midpointgradientcell{9.8}{2.6}{11.1}{0}{neg}{pos}{\opacity}{0} & \midpointgradientcell{-1.6}{-11.9}{0.3}{0}{neg}{pos}{\opacity}{0} & \midpointgradientcell{-3.5}{-8.8}{-0.6}{0}{neg}{pos}{\opacity}{0} & \midpointgradientcell{4.1}{0.8}{8.4}{0}{neg}{pos}{\opacity}{0} & \midpointgradientcell{0.8}{-8.6}{0.9}{0}{neg}{pos}{\opacity}{1} & \midpointgradientcell{-4.6}{-10.5}{-1.1}{0}{neg}{pos}{\opacity}{0} & \midpointgradientcell{-3.5}{-16.0}{3.0}{0}{neg}{pos}{\opacity}{0} & \midpointgradientcell{22.6}{-11.1}{22.8}{0}{neg}{pos}{\opacity}{0} & \midpointgradientcell{8.4}{-26.9}{8.5}{0}{neg}{pos}{\opacity}{1} \\
 & Replace & \midpointgradientcell{8.3}{2.6}{11.1}{0}{neg}{pos}{\opacity}{0} & \midpointgradientcell{-7.3}{-11.9}{0.3}{0}{neg}{pos}{\opacity}{0} & \midpointgradientcell{-7.9}{-8.8}{-0.6}{0}{neg}{pos}{\opacity}{0} & \midpointgradientcell{7.2}{0.8}{8.4}{0}{neg}{pos}{\opacity}{0} & \midpointgradientcell{-3.2}{-8.6}{0.9}{0}{neg}{pos}{\opacity}{0} & \midpointgradientcell{-8.8}{-10.5}{-1.1}{0}{neg}{pos}{\opacity}{0} & \midpointgradientcell{-15.9}{-16.0}{3.0}{0}{neg}{pos}{\opacity}{0} & \midpointgradientcell{19.3}{-11.1}{22.8}{0}{neg}{pos}{\opacity}{0} & \midpointgradientcell{3.3}{-26.9}{8.5}{0}{neg}{pos}{\opacity}{0} \\
\midrule
\parbox[t]{2mm}{\multirow{3}{*}{\rotatebox[origin=c]{90}{
 \parbox[c]{0.5cm}{\centering \scriptsize Doc }}}} 
 & Prepend & \midpointgradientcell{8.5}{2.6}{11.1}{0}{neg}{pos}{\opacity}{0} & \midpointgradientcell{-2.2}{-11.9}{0.3}{0}{neg}{pos}{\opacity}{0} & \midpointgradientcell{-1.9}{-8.8}{-0.6}{0}{neg}{pos}{\opacity}{0} & \midpointgradientcell{5.9}{0.8}{8.4}{0}{neg}{pos}{\opacity}{0} & \midpointgradientcell{-2.0}{-8.6}{0.9}{0}{neg}{pos}{\opacity}{0} & \midpointgradientcell{-3.1}{-10.5}{-1.1}{0}{neg}{pos}{\opacity}{0} & \midpointgradientcell{1.4}{-16.0}{3.0}{0}{neg}{pos}{\opacity}{0} & \midpointgradientcell{-5.4}{-11.1}{22.8}{0}{neg}{pos}{\opacity}{0} & \midpointgradientcell{-12.4}{-26.9}{8.5}{0}{neg}{pos}{\opacity}{0} \\
 & Append & \midpointgradientcell{10.3}{2.6}{11.1}{0}{neg}{pos}{\opacity}{0} & \midpointgradientcell{-0.8}{-11.9}{0.3}{0}{neg}{pos}{\opacity}{0} & \midpointgradientcell{-1.4}{-8.8}{-0.6}{0}{neg}{pos}{\opacity}{0} & \midpointgradientcell{4.0}{0.8}{8.4}{0}{neg}{pos}{\opacity}{0} & \midpointgradientcell{-1.4}{-8.6}{0.9}{0}{neg}{pos}{\opacity}{0} & \midpointgradientcell{-2.2}{-10.5}{-1.1}{0}{neg}{pos}{\opacity}{0} & \midpointgradientcell{0.4}{-16.0}{3.0}{0}{neg}{pos}{\opacity}{0} & \midpointgradientcell{-6.8}{-11.1}{22.8}{0}{neg}{pos}{\opacity}{0} & \midpointgradientcell{-8.6}{-26.9}{8.5}{0}{neg}{pos}{\opacity}{0} \\
 & Replace & \midpointgradientcell{9.3}{2.6}{11.1}{0}{neg}{pos}{\opacity}{0} & \midpointgradientcell{-8.9}{-11.9}{0.3}{0}{neg}{pos}{\opacity}{0} & \midpointgradientcell{-6.2}{-8.8}{-0.6}{0}{neg}{pos}{\opacity}{0} & \midpointgradientcell{8.3}{0.8}{8.4}{0}{neg}{pos}{\opacity}{1} & \midpointgradientcell{-6.9}{-8.6}{0.9}{0}{neg}{pos}{\opacity}{0} & \midpointgradientcell{-8.8}{-10.5}{-1.1}{0}{neg}{pos}{\opacity}{0} & \midpointgradientcell{-4.1}{-16.0}{3.0}{0}{neg}{pos}{\opacity}{0} & \midpointgradientcell{-11.0}{-11.1}{22.8}{0}{neg}{pos}{\opacity}{0} & \midpointgradientcell{-20.1}{-26.9}{8.5}{0}{neg}{pos}{\opacity}{0} \\
\midrule
\parbox[t]{2mm}{\multirow{9}{*}{\rotatebox[origin=c]{90}{
 \parbox[c]{1cm}{\centering \scriptsize Both }}}} 
 & Prepend/Prepend & \midpointgradientcell{9.4}{2.6}{11.1}{0}{neg}{pos}{\opacity}{0} & \midpointgradientcell{-2.2}{-11.9}{0.3}{0}{neg}{pos}{\opacity}{0} & \midpointgradientcell{-2.0}{-8.8}{-0.6}{0}{neg}{pos}{\opacity}{0} & \midpointgradientcell{5.9}{0.8}{8.4}{0}{neg}{pos}{\opacity}{0} & \midpointgradientcell{-4.0}{-8.6}{0.9}{0}{neg}{pos}{\opacity}{0} & \midpointgradientcell{-4.6}{-10.5}{-1.1}{0}{neg}{pos}{\opacity}{0} & \midpointgradientcell{1.5}{-16.0}{3.0}{0}{neg}{pos}{\opacity}{0} & \midpointgradientcell{-9.7}{-11.1}{22.8}{0}{neg}{pos}{\opacity}{0} & \midpointgradientcell{-19.8}{-26.9}{8.5}{0}{neg}{pos}{\opacity}{0} \\
 & Prepend/Append & \midpointgradientcell{11.0}{2.6}{11.1}{0}{neg}{pos}{\opacity}{1} & \midpointgradientcell{-0.9}{-11.9}{0.3}{0}{neg}{pos}{\opacity}{0} & \midpointgradientcell{-1.9}{-8.8}{-0.6}{0}{neg}{pos}{\opacity}{0} & \midpointgradientcell{4.1}{0.8}{8.4}{0}{neg}{pos}{\opacity}{0} & \midpointgradientcell{-3.3}{-8.6}{0.9}{0}{neg}{pos}{\opacity}{0} & \midpointgradientcell{-2.8}{-10.5}{-1.1}{0}{neg}{pos}{\opacity}{0} & \midpointgradientcell{0.5}{-16.0}{3.0}{0}{neg}{pos}{\opacity}{0} & \midpointgradientcell{-8.7}{-11.1}{22.8}{0}{neg}{pos}{\opacity}{0} & \midpointgradientcell{-18.3}{-26.9}{8.5}{0}{neg}{pos}{\opacity}{0} \\
 & Prepend/Replace & \midpointgradientcell{9.6}{2.6}{11.1}{0}{neg}{pos}{\opacity}{0} & \midpointgradientcell{-9.0}{-11.9}{0.3}{0}{neg}{pos}{\opacity}{0} & \midpointgradientcell{-6.2}{-8.8}{-0.6}{0}{neg}{pos}{\opacity}{0} & \midpointgradientcell{8.1}{0.8}{8.4}{0}{neg}{pos}{\opacity}{0} & \midpointgradientcell{-8.5}{-8.6}{0.9}{0}{neg}{pos}{\opacity}{0} & \midpointgradientcell{-9.3}{-10.5}{-1.1}{0}{neg}{pos}{\opacity}{0} & \midpointgradientcell{-5.1}{-16.0}{3.0}{0}{neg}{pos}{\opacity}{0} & \midpointgradientcell{-10.0}{-11.1}{22.8}{0}{neg}{pos}{\opacity}{0} & \midpointgradientcell{-26.8}{-26.9}{8.5}{0}{neg}{pos}{\opacity}{0} \\
 & Append/Prenpend & \midpointgradientcell{3.5}{2.6}{11.1}{0}{neg}{pos}{\opacity}{0} & \midpointgradientcell{-2.0}{-11.9}{0.3}{0}{neg}{pos}{\opacity}{0} & \midpointgradientcell{-2.2}{-8.8}{-0.6}{0}{neg}{pos}{\opacity}{0} & \midpointgradientcell{3.6}{0.8}{8.4}{0}{neg}{pos}{\opacity}{0} & \midpointgradientcell{0.1}{-8.6}{0.9}{0}{neg}{pos}{\opacity}{0} & \midpointgradientcell{-3.8}{-10.5}{-1.1}{0}{neg}{pos}{\opacity}{0} & \midpointgradientcell{-0.1}{-16.0}{3.0}{0}{neg}{pos}{\opacity}{0} & \midpointgradientcell{22.7}{-11.1}{22.8}{0}{neg}{pos}{\opacity}{1} & \midpointgradientcell{8.3}{-26.9}{8.5}{0}{neg}{pos}{\opacity}{0} \\
 & Append/Append & \midpointgradientcell{2.7}{2.6}{11.1}{0}{neg}{pos}{\opacity}{0} & \midpointgradientcell{-1.7}{-11.9}{0.3}{0}{neg}{pos}{\opacity}{0} & \midpointgradientcell{-1.1}{-8.8}{-0.6}{0}{neg}{pos}{\opacity}{0} & \midpointgradientcell{4.8}{0.8}{8.4}{0}{neg}{pos}{\opacity}{0} & \midpointgradientcell{-3.5}{-8.6}{0.9}{0}{neg}{pos}{\opacity}{0} & \midpointgradientcell{-2.0}{-10.5}{-1.1}{0}{neg}{pos}{\opacity}{0} & \midpointgradientcell{-0.5}{-16.0}{3.0}{0}{neg}{pos}{\opacity}{0} & \midpointgradientcell{-5.3}{-11.1}{22.8}{0}{neg}{pos}{\opacity}{0} & \midpointgradientcell{-9.0}{-26.9}{8.5}{0}{neg}{pos}{\opacity}{0} \\
 & Append/Replace & \midpointgradientcell{3.0}{2.6}{11.1}{0}{neg}{pos}{\opacity}{0} & \midpointgradientcell{-1.7}{-11.9}{0.3}{0}{neg}{pos}{\opacity}{0} & \midpointgradientcell{-1.3}{-8.8}{-0.6}{0}{neg}{pos}{\opacity}{0} & \midpointgradientcell{4.6}{0.8}{8.4}{0}{neg}{pos}{\opacity}{0} & \midpointgradientcell{-5.6}{-8.6}{0.9}{0}{neg}{pos}{\opacity}{0} & \midpointgradientcell{-2.2}{-10.5}{-1.1}{0}{neg}{pos}{\opacity}{0} & \midpointgradientcell{-0.3}{-16.0}{3.0}{0}{neg}{pos}{\opacity}{0} & \midpointgradientcell{-8.0}{-11.1}{22.8}{0}{neg}{pos}{\opacity}{0} & \midpointgradientcell{-18.8}{-26.9}{8.5}{0}{neg}{pos}{\opacity}{0} \\
 & Replace/Prepend & \midpointgradientcell{4.0}{2.6}{11.1}{0}{neg}{pos}{\opacity}{0} & \midpointgradientcell{-2.8}{-11.9}{0.3}{0}{neg}{pos}{\opacity}{0} & \midpointgradientcell{-1.2}{-8.8}{-0.6}{0}{neg}{pos}{\opacity}{0} & \midpointgradientcell{1.6}{0.8}{8.4}{0}{neg}{pos}{\opacity}{0} & \midpointgradientcell{-0.6}{-8.6}{0.9}{0}{neg}{pos}{\opacity}{0} & \midpointgradientcell{-3.2}{-10.5}{-1.1}{0}{neg}{pos}{\opacity}{0} & \midpointgradientcell{2.9}{-16.0}{3.0}{0}{neg}{pos}{\opacity}{1} & \midpointgradientcell{-3.0}{-11.1}{22.8}{0}{neg}{pos}{\opacity}{0} & \midpointgradientcell{-2.1}{-26.9}{8.5}{0}{neg}{pos}{\opacity}{0} \\
 & Replace/Append & \midpointgradientcell{5.9}{2.6}{11.1}{0}{neg}{pos}{\opacity}{0} & \midpointgradientcell{0.2}{-11.9}{0.3}{0}{neg}{pos}{\opacity}{1} & \midpointgradientcell{-0.7}{-8.8}{-0.6}{0}{neg}{pos}{\opacity}{1} & \midpointgradientcell{0.9}{0.8}{8.4}{0}{neg}{pos}{\opacity}{0} & \midpointgradientcell{0.6}{-8.6}{0.9}{0}{neg}{pos}{\opacity}{0} & \midpointgradientcell{-1.2}{-10.5}{-1.1}{0}{neg}{pos}{\opacity}{1} & \midpointgradientcell{1.2}{-16.0}{3.0}{0}{neg}{pos}{\opacity}{0} & \midpointgradientcell{-1.5}{-11.1}{22.8}{0}{neg}{pos}{\opacity}{0} & \midpointgradientcell{-0.9}{-26.9}{8.5}{0}{neg}{pos}{\opacity}{0} \\
 & Replace/Replace & \midpointgradientcell{5.7}{2.6}{11.1}{0}{neg}{pos}{\opacity}{0} & \midpointgradientcell{-11.8}{-11.9}{0.3}{0}{neg}{pos}{\opacity}{0} & \midpointgradientcell{-8.7}{-8.8}{-0.6}{0}{neg}{pos}{\opacity}{0} & \midpointgradientcell{4.4}{0.8}{8.4}{0}{neg}{pos}{\opacity}{0} & \midpointgradientcell{-5.3}{-8.6}{0.9}{0}{neg}{pos}{\opacity}{0} & \midpointgradientcell{-10.4}{-10.5}{-1.1}{0}{neg}{pos}{\opacity}{0} & \midpointgradientcell{-1.0}{-16.0}{3.0}{0}{neg}{pos}{\opacity}{0} & \midpointgradientcell{-5.0}{-11.1}{22.8}{0}{neg}{pos}{\opacity}{0} & \midpointgradientcell{-9.1}{-26.9}{8.5}{0}{neg}{pos}{\opacity}{0} \\
\bottomrule
\end{tabular}}
\caption{How different placements of the expansions affect results (e.g. prepend/append/replace). Colors indicate a \hlpos{positive} or \hlneg{negative} delta over \hlbase{scores for no expansion}. Although there are small differences \textbf{the overall trends are the same}.\vspace{-0.5em}}
\label{tab:placement}
\end{table*}

\section{Number of Documents Reranked}
\label{app:rerank_size}
All experiments reranked the top 100 documents, as described in the main paper. However, here we show an ablation where we increase the number of reranked documents to include the top 1000 or 10,000 documents. We see that there is little change in how expansions affect the models.
\begin{table*}[t!]
\centering
\small
\resizebox{\textwidth}{!}{\begin{tabular}{ll|ccc|ccc|ccc}
\toprule
\multicolumn{2}{l}{} & \multicolumn{3}{c|}{\textbf{
TREC DL 2019}} & \multicolumn{3}{c}{\textbf{NFCorpus}} & \multicolumn{3}{c}{\textbf{Arguana}} \\
\scriptsize \textbf{Type} & \scriptsize\textbf{Model} & \scriptsize DPR  & \scriptsize Contriever FT & \scriptsize MonoT5-3B  & \scriptsize DPR  & \scriptsize Contriever FT & \scriptsize MonoT5-3B & \scriptsize DPR  & \scriptsize Contriever FT & \scriptsize MonoT5-3B \\
\midrule
 \parbox[t]{2mm}{\multirow{4}{*}{\rotatebox[origin=c]{90}{
 \parbox[c]{1cm}{\centering \scriptsize 100 Docs }}}}  & \textit{No Expansion} & \gradientcell{38.4}{24.0}{71.8}{low}{high}{\opacity}{0} & \gradientcell{62.3}{24.0}{71.8}{low}{high}{\opacity}{0} & \gradientcell{71.7}{24.0}{71.8}{low}{high}{\opacity}{0} & \gradientcell{24.1}{24.0}{71.8}{low}{high}{\opacity}{0} & \gradientcell{34.6}{24.0}{71.8}{low}{high}{\opacity}{0} & \gradientcell{39.2}{24.0}{71.8}{low}{high}{\opacity}{0} & \gradientcell{34.9}{24.0}{71.8}{low}{high}{\opacity}{0} & \gradientcell{48.8}{24.0}{71.8}{low}{high}{\opacity}{0} & \gradientcell{42.4}{24.0}{71.8}{low}{high}{\opacity}{0} \\

 & Q-LM PRF & \midpointgradientcell{6.6}{3.0}{10.9}{0}{neg}{pos}{\opacity}{0} & \midpointgradientcell{1.6}{-0.3}{1.7}{0}{neg}{pos}{\opacity}{1} & \midpointgradientcell{-2.8}{-5.1}{-1.9}{0}{neg}{pos}{\opacity}{0} & \midpointgradientcell{0.2}{0.1}{1.5}{0}{neg}{pos}{\opacity}{0} & \midpointgradientcell{-0.4}{-0.5}{0.7}{0}{neg}{pos}{\opacity}{0} & \midpointgradientcell{-2.8}{-3.1}{-0.4}{0}{neg}{pos}{\opacity}{0} & \midpointgradientcell{-7.1}{-7.2}{1.7}{0}{neg}{pos}{\opacity}{0} & \midpointgradientcell{-3.6}{-3.7}{2.1}{0}{neg}{pos}{\opacity}{0} & \midpointgradientcell{6.8}{-2.0}{6.9}{0}{neg}{pos}{\opacity}{1} \\
 & D2Q & \midpointgradientcell{3.1}{3.0}{10.9}{0}{neg}{pos}{\opacity}{0} & \midpointgradientcell{-0.2}{-0.3}{1.7}{0}{neg}{pos}{\opacity}{0} & \midpointgradientcell{-2.0}{-5.1}{-1.9}{0}{neg}{pos}{\opacity}{1} & \midpointgradientcell{1.3}{0.1}{1.5}{0}{neg}{pos}{\opacity}{0} & \midpointgradientcell{0.6}{-0.5}{0.7}{0}{neg}{pos}{\opacity}{1} & \midpointgradientcell{-0.5}{-3.1}{-0.4}{0}{neg}{pos}{\opacity}{1} & \midpointgradientcell{1.6}{-7.2}{1.7}{0}{neg}{pos}{\opacity}{1} & \midpointgradientcell{2.0}{-3.7}{2.1}{0}{neg}{pos}{\opacity}{1} & \midpointgradientcell{-1.9}{-2.0}{6.9}{0}{neg}{pos}{\opacity}{0} \\
 & Q-LM PRF + D2Q & \midpointgradientcell{10.8}{3.0}{10.9}{0}{neg}{pos}{\opacity}{1} & \midpointgradientcell{0.6}{-0.3}{1.7}{0}{neg}{pos}{\opacity}{0} & \midpointgradientcell{-5.0}{-5.1}{-1.9}{0}{neg}{pos}{\opacity}{0} & \midpointgradientcell{1.4}{0.1}{1.5}{0}{neg}{pos}{\opacity}{1} & \midpointgradientcell{-0.1}{-0.5}{0.7}{0}{neg}{pos}{\opacity}{0} & \midpointgradientcell{-3.0}{-3.1}{-0.4}{0}{neg}{pos}{\opacity}{0} & \midpointgradientcell{-4.4}{-7.2}{1.7}{0}{neg}{pos}{\opacity}{0} & \midpointgradientcell{-2.5}{-3.7}{2.1}{0}{neg}{pos}{\opacity}{0} & \midpointgradientcell{5.2}{-2.0}{6.9}{0}{neg}{pos}{\opacity}{0} \\

\midrule

 \parbox[t]{2mm}{\multirow{4}{*}{\rotatebox[origin=c]{90}{
 \parbox[c]{1cm}{\centering \scriptsize 1k docs }}}}  & \textit{No Expansion} & \gradientcell{29.2}{21.4}{72.7}{low}{high}{\opacity}{0} & \gradientcell{64.6}{21.4}{72.7}{low}{high}{\opacity}{0} & \gradientcell{72.6}{21.4}{72.7}{low}{high}{\opacity}{0} & \gradientcell{21.5}{21.4}{72.7}{low}{high}{\opacity}{0} & \gradientcell{34.2}{21.4}{72.7}{low}{high}{\opacity}{0} & \gradientcell{40.0}{21.4}{72.7}{low}{high}{\opacity}{0} & \gradientcell{29.5}{21.4}{72.7}{low}{high}{\opacity}{0} & \gradientcell{48.7}{21.4}{72.7}{low}{high}{\opacity}{0} & \gradientcell{38.0}{21.4}{72.7}{low}{high}{\opacity}{0} \\

 & PRF & \midpointgradientcell{4.1}{4.0}{11.0}{0}{neg}{pos}{\opacity}{0} & \midpointgradientcell{-0.6}{-0.7}{0.8}{0}{neg}{pos}{\opacity}{0} & \midpointgradientcell{-4.2}{-6.4}{-1.6}{0}{neg}{pos}{\opacity}{0} & \midpointgradientcell{-0.9}{-1.0}{1.2}{0}{neg}{pos}{\opacity}{0} & \midpointgradientcell{0.1}{0.0}{0.8}{0}{neg}{pos}{\opacity}{0} & \midpointgradientcell{-3.5}{-4.0}{-0.2}{0}{neg}{pos}{\opacity}{0} & \midpointgradientcell{-10.1}{-10.2}{-5.1}{0}{neg}{pos}{\opacity}{0} & \midpointgradientcell{-14.5}{-14.6}{-11.2}{0}{neg}{pos}{\opacity}{0} & \midpointgradientcell{-1.9}{-10.5}{-1.8}{0}{neg}{pos}{\opacity}{1} \\
 & D2Q & \midpointgradientcell{5.2}{4.0}{11.0}{0}{neg}{pos}{\opacity}{0} & \midpointgradientcell{0.7}{-0.7}{0.8}{0}{neg}{pos}{\opacity}{1} & \midpointgradientcell{-1.7}{-6.4}{-1.6}{0}{neg}{pos}{\opacity}{1} & \midpointgradientcell{1.1}{-1.0}{1.2}{0}{neg}{pos}{\opacity}{1} & \midpointgradientcell{0.7}{0.0}{0.8}{0}{neg}{pos}{\opacity}{1} & \midpointgradientcell{-0.3}{-4.0}{-0.2}{0}{neg}{pos}{\opacity}{1} & \midpointgradientcell{-5.2}{-10.2}{-5.1}{0}{neg}{pos}{\opacity}{1} & \midpointgradientcell{-11.3}{-14.6}{-11.2}{0}{neg}{pos}{\opacity}{1} & \midpointgradientcell{-10.4}{-10.5}{-1.8}{0}{neg}{pos}{\opacity}{0} \\
 & PRF + D2Q & \midpointgradientcell{10.9}{4.0}{11.0}{0}{neg}{pos}{\opacity}{1} & \midpointgradientcell{-0.1}{-0.7}{0.8}{0}{neg}{pos}{\opacity}{0} & \midpointgradientcell{-6.3}{-6.4}{-1.6}{0}{neg}{pos}{\opacity}{0} & \midpointgradientcell{0.7}{-1.0}{1.2}{0}{neg}{pos}{\opacity}{0} & \midpointgradientcell{0.6}{0.0}{0.8}{0}{neg}{pos}{\opacity}{0} & \midpointgradientcell{-3.9}{-4.0}{-0.2}{0}{neg}{pos}{\opacity}{0} & \midpointgradientcell{-8.8}{-10.2}{-5.1}{0}{neg}{pos}{\opacity}{0} & \midpointgradientcell{-13.5}{-14.6}{-11.2}{0}{neg}{pos}{\opacity}{0} & \midpointgradientcell{-2.8}{-10.5}{-1.8}{0}{neg}{pos}{\opacity}{0} \\
 
 \midrule
  \parbox[t]{2mm}{\multirow{4}{*}{\rotatebox[origin=c]{90}{
 \parbox[c]{1cm}{\centering \scriptsize 10k docs  }}}}  & \textit{No Expansion} & \gradientcell{28.5}{21.4}{73.2}{low}{high}{\opacity}{0} & \gradientcell{64.6}{21.4}{73.2}{low}{high}{\opacity}{0} & \gradientcell{73.1}{21.4}{73.2}{low}{high}{\opacity}{0} & \gradientcell{21.5}{21.4}{73.2}{low}{high}{\opacity}{0} & \gradientcell{34.2}{21.4}{73.2}{low}{high}{\opacity}{0} & \gradientcell{40.0}{21.4}{73.2}{low}{high}{\opacity}{0} & \gradientcell{23.5}{21.4}{73.2}{low}{high}{\opacity}{0} & \gradientcell{36.2}{21.4}{73.2}{low}{high}{\opacity}{0} & \gradientcell{29.7}{21.4}{73.2}{low}{high}{\opacity}{0} \\

 & PRF & \midpointgradientcell{3.4}{3.3}{10.2}{0}{neg}{pos}{\opacity}{0} & \midpointgradientcell{0.6}{-1.5}{2.9}{0}{neg}{pos}{\opacity}{0} & \midpointgradientcell{-4.3}{-4.4}{-1.4}{0}{neg}{pos}{\opacity}{0} & \midpointgradientcell{-0.7}{-0.8}{1.0}{0}{neg}{pos}{\opacity}{0} & \midpointgradientcell{0.6}{0.4}{0.7}{0}{neg}{pos}{\opacity}{1} & \midpointgradientcell{-3.2}{-3.5}{-0.1}{0}{neg}{pos}{\opacity}{0} & \midpointgradientcell{-3.9}{-4.0}{1.3}{0}{neg}{pos}{\opacity}{0} & \midpointgradientcell{-1.9}{-2.0}{1.7}{0}{neg}{pos}{\opacity}{0} & \midpointgradientcell{6.6}{-2.3}{6.7}{0}{neg}{pos}{\opacity}{1} \\
 & D2Q & \midpointgradientcell{4.2}{3.3}{10.2}{0}{neg}{pos}{\opacity}{0} & \midpointgradientcell{-1.4}{-1.5}{2.9}{0}{neg}{pos}{\opacity}{0} & \midpointgradientcell{-1.5}{-4.4}{-1.4}{0}{neg}{pos}{\opacity}{1} & \midpointgradientcell{0.9}{-0.8}{1.0}{0}{neg}{pos}{\opacity}{1} & \midpointgradientcell{0.5}{0.4}{0.7}{0}{neg}{pos}{\opacity}{0} & \midpointgradientcell{-0.2}{-3.5}{-0.1}{0}{neg}{pos}{\opacity}{1} & \midpointgradientcell{1.2}{-4.0}{1.3}{0}{neg}{pos}{\opacity}{1} & \midpointgradientcell{1.6}{-2.0}{1.7}{0}{neg}{pos}{\opacity}{1} & \midpointgradientcell{-2.2}{-2.3}{6.7}{0}{neg}{pos}{\opacity}{0} \\
 & PRF + D2Q & \midpointgradientcell{10.1}{3.3}{10.2}{0}{neg}{pos}{\opacity}{1} & \midpointgradientcell{2.8}{-1.5}{2.9}{0}{neg}{pos}{\opacity}{1} & \midpointgradientcell{-4.2}{-4.4}{-1.4}{0}{neg}{pos}{\opacity}{0} & \midpointgradientcell{0.3}{-0.8}{1.0}{0}{neg}{pos}{\opacity}{0} & \midpointgradientcell{0.6}{0.4}{0.7}{0}{neg}{pos}{\opacity}{1} & \midpointgradientcell{-3.4}{-3.5}{-0.1}{0}{neg}{pos}{\opacity}{0} & \midpointgradientcell{-2.1}{-4.0}{1.3}{0}{neg}{pos}{\opacity}{0} & \midpointgradientcell{-1.0}{-2.0}{1.7}{0}{neg}{pos}{\opacity}{0} & \midpointgradientcell{5.8}{-2.3}{6.7}{0}{neg}{pos}{\opacity}{0} \\
 
\bottomrule
\end{tabular}}
\caption{How different numbers of documents used for reranking affect the results. Colors indicate a \hlpos{positive} or \hlneg{negative} delta over \hlbase{scores for no expansion}. Although there are small differences between models \textbf{the overall trends are the same}. Note that NFCorpus's collection is roughly 3k examples and so changing from 1k to 10k reranking makes little difference.\vspace{-0.5em}}
\label{tab:rerank_size}
\end{table*}

\section{Example Input/Output for expansions}
Here we show example input and output for expansion for query and document expansions.
\label{app:examples}

\subsection{Document Expansion}
We show example expansion from this document from MS MARCO: 
\begin{tcolorbox}
Whale Shark. The Whale Shark is the largest and the heaviest living shark because its weight is more than 21 tons and its length is more than 12 meter. It is found in open oceans and warm waters. It feeds on plankton but some times it also uses other small fishes as food.
\end{tcolorbox}

\subsubsection{Doc2Query}
\begin{tcolorbox}
What is the largest and heaviest living shark in the world? 

What are the characteristics of the Whale Shark? 

Where can the Whale Shark be found? 

What does the Whale Shark feed on? 

What is the size and weight of the Whale Shark? 
\end{tcolorbox}

\subsubsection{D-LM PRF}
\begin{tcolorbox}
The Whale Shark is the largest and heaviest living shark, weighing over 21 tons and measuring over 12 meters in length. It is found in open oceans and warm waters. While it primarily feeds on plankton, it occasionally consumes other small fishes as well.
\end{tcolorbox}

\subsection{Query Expansion}

For the MS MARCO query ``why did the us volunterilay enter ww1" the following output was generated. For prompt formatting, see \autoref{app:prompts}.

\subsubsection{HyDE}
\begin{tcolorbox}
The United States voluntarily entered World War I primarily due to a combination of factors. Firstly, the sinking of the British ocean liner RMS Lusitania by a German U-boat in 1915, resulting in the loss of American lives, created a strong public outcry for intervention. Additionally, Germany's resumption of unrestricted submarine warfare in 1917 further escalated tensions. Furthermore, the interception of the Zimmermann Telegram, which revealed Germany's proposal to Mexico for a military alliance against the United States, heightened fears of potential threats to American security. These events, coupled with a desire to protect American interests and uphold democratic principles, ultimately led the United States to voluntarily join the war effort.
\end{tcolorbox}

\subsubsection{Q-LM PRF}
\begin{tcolorbox}
What were the factors leading to the United States' voluntary entry into WW1 and what were the consequences of their decision to join the Allies?
\end{tcolorbox}

\subsubsection{Chain of Thought}
\begin{tcolorbox}
The United States voluntarily entered World War I primarily due to a combination of factors. One of the key reasons was the German unrestricted submarine warfare, which threatened American lives and interests at sea. Additionally, the interception and decoding of the Zimmerman Telegram, in which Germany proposed an alliance with Mexico against the US, further fueled public sentiment for intervention. Moreover, there was a desire to protect American economic interests, as the US had significant trade ties with the Allied powers. Lastly, President Woodrow Wilson's belief in promoting democracy and his vision for a post-war world order based on the principles of self-determination and collective security also played a role in the US decision to enter the war.
\end{tcolorbox}

\section{Prompts for LM expansions}
\label{app:prompts}
Here we show the prompts used for LM query and document expansion.

\subsection{Query Expansions}
\subsubsection{HyDE}
We use the prompts from their paper, which are dataset specific, in this manner:
\begin{tcolorbox}
<SPECIFIC PROMPT> \\

<Type of Input>: QUERY \\
<TYPE OF OUTPUT> (one short paragraph max):
\end{tcolorbox}

for example on Clinical Trials we use:
\begin{tcolorbox}
Please write a clinical trial summary that would apply to the following patient. \\

Patient Info: QUERY \\
Trial Summary (one short paragraph max):
\end{tcolorbox}

and on FiQA we use:
\begin{tcolorbox}
Please write a financial article passage to answer the question \\

Question: QUERY \\
Passage (one short paragraph max):
\end{tcolorbox}

\subsubsection{Q-LM PRF}
\begin{tcolorbox}
You are a query expansion engine, primed and ready to take in text and output additional keywords will provide new and expanded context behind the original input. Your extensive world knowledge and linguistic creativity enables you to provide questions that maximally optimize the new questions to find new websites. You **always** provide creative synonyms and acronym expansions in your new queries that will provide additional insight. \\

Be sure to use new words and spell out acronyms (or add new acronyms). Hint: think of ***new synonyms and/or acronyms*** for ``QUESTION" using these documents for inspiration: \\

DOCUMENTS \\

Return the following information, filling it in:

Input: QUESTION 

Comma Separated List of 10 important New Keywords: ``````NEW KEYWORDS HERE""" 

New Question (combining Input and New Keywords, only **one** new question that expands upon the Input): ``````NEW QUESTION HERE""" \\ \\

Your output:
\end{tcolorbox}

\subsubsection{Chain of Thought}
We use a the same specific prompt for CoT as we do for HyDE. The format is as follows: 
\begin{tcolorbox}
<SPECIFIC PROMPT> \\

QUESTION \\

Give the rationale (one short paragraph max) before answering.
\end{tcolorbox}

\subsection{Document Expansions}

\begin{figure*}[h]
    \centering       \includegraphics[width=0.49\columnwidth,trim=0cm 0cm 0.0cm -0.25cm]{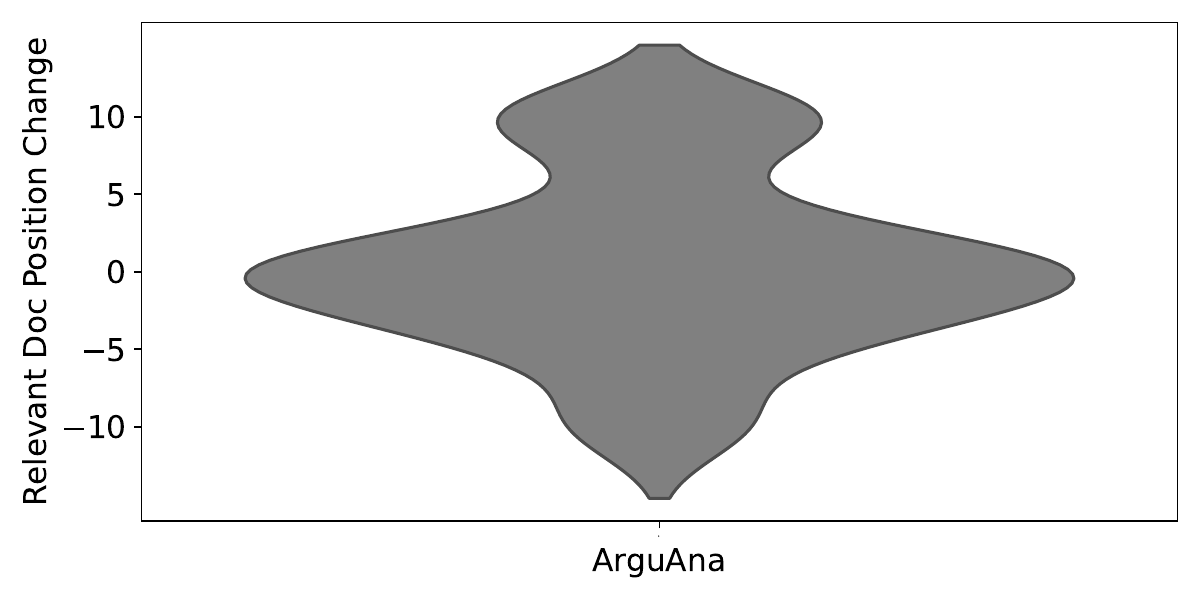}
       \includegraphics[width=0.49\columnwidth,trim=0cm 0cm 0.0cm -0.25cm]{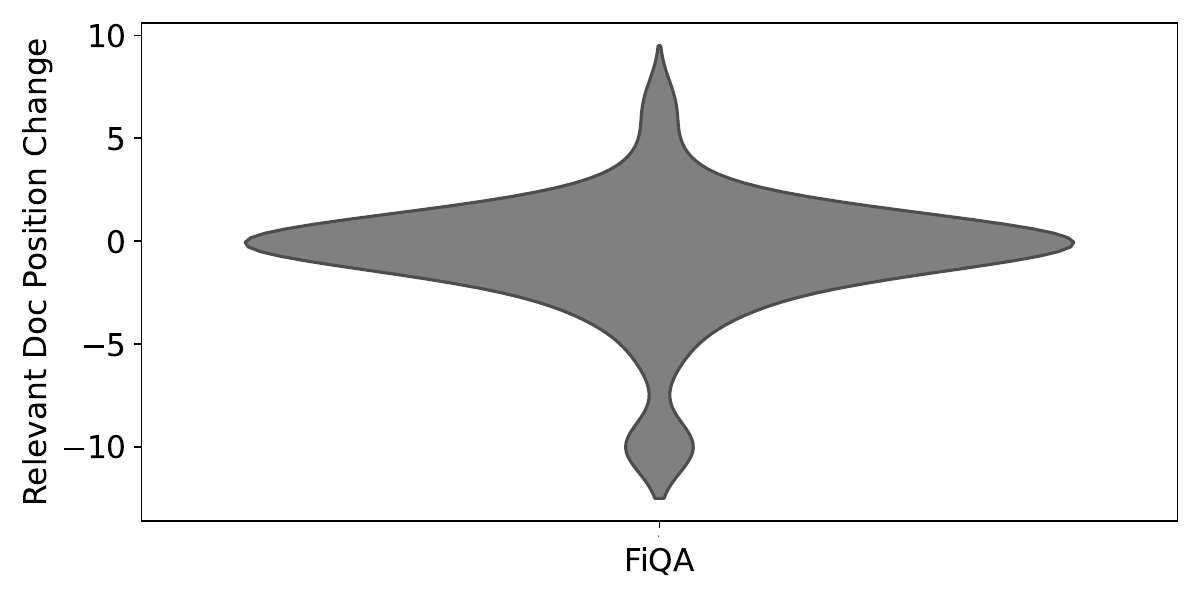}

    \caption{Number of positions relevant documents change when using expansion. Negative values indicate the document was ranked lower. Results are similar to TREC DL 2019 for FiQA which shows lowered nDCG while for Arguana nDCG scores increase as seen by the change in positions being positive.\vspace{-1.0em}}
    \label{fig:position_change_app}
\end{figure*}

\subsubsection{D-LM PRF}
\begin{tcolorbox}
Change the following document to answer these questions, if they are partially answered by the document. If the queries are not relevant, ignore them. Your new documents should be one concise paragraph following the examples. \\
 
Example 1: \\

Queries:

1. ``how much caffeine is in a 12 ounce cup of coffee?"

2. ``what are the effects of alcohol and caffeine"

3. ``what can pregnant women not do?"

Document: ``We don't know a lot about the effects of caffeine during pregnancy on you and your baby. So it's best to limit the amount you get each day. If you are pregnant, limit caffeine to 200 milligrams each day. This is about the amount in 1½ 8-ounce cups of coffee or one 12-ounce cup of coffee." 

New Document (similar to Document): ``There is a lack of research about the effects of caffeine during pregnancy on you and your baby. So it's best to limit the amount you get each day. If you are pregnant, limit caffeine to 200 milligrams (mg) each day. This is about the amount in 1½ 8-ounce cups of coffee or one 12-ounce cup of coffee (e.g. 200 milligrams)." \\

Example 2: \\

Queries:

QUERIES

Document: ``DOCUMENT"

New Document (similar to Document): 
\end{tcolorbox}

\subsubsection{Doc2Query}
\begin{tcolorbox}
You are an optimized query expansion model, ExpansionGPT. You will write 5 queries for the given document that help retrieval models better find this document during search. \\

Document: ``QUESTION" \\

Queries:

\end{tcolorbox}

\end{document}